\documentclass[aip,jcp,showpacs,12pt]{revtex4-1}
\usepackage{graphicx}
\usepackage{epsfig}
\usepackage{dcolumn}
\usepackage{bm}
\usepackage{color}
\usepackage{ulem}
\usepackage{float}

\begin{document}

\title{Layer-by-layer assembly of patchy particles as a route to non-trivial structures}
\author{Niladri Patra}
\author{Alexei V. Tkachenko}
\affiliation{Center for Functional Nanomaterials, Brookhaven National Laboratory, Upton, New York 11973}
\email{oleksiyt@bnl.gov}

\begin{abstract}
We propose a new strategy  for robust high-quality  self-assembly of non-trivial periodic structures out of patchy particles, and investigate it with  Brownian Dynamics (BD) simulations. Its  first element is the use of  specific patch-patch and shell-shell  interactions between the particles, that can be implemented  through differential functionalization of patched and shell regions with specific  DNA strands.  The other  key element of our approach is the use of  layer-by-layer  protocol that allows  to avoid a formations of undesired random aggregates. As an example, we design and self-assemble ``in silico"  a version of a Double Diamond (DD) lattice in which four particle types are arranged into BCC crystal made of four FCC sub-lattices.   The  lattice can be further converted to Cubic Diamond (CD) by selective removal of the  particles of certain types. Our results demonstrate that by combining the directionality, selectivity of interactions and the layer-by-layer protocol, a   high quality robust self-assembly can be achieved.  
\end{abstract}
\maketitle

\section{Introduction}

In recent years, the field of nanoparticle (NP)  and colloidal self-assembly had demonstrated  an impressive progress. It was marked by an  emergence of new   approaches, such as the  use of anisotropic particles and specific key-lock interactions. Among the most popular model systems with anisotropic potentials are so-called patchy particles, i.e. colloids with chemically distinct ``patches" on their surfaces \cite{Glo04,Patchy_cluster,Kern,Scior_Nat_phys,Patchy_fab,Pine_patchy_2003}. The  key-lock interactions are typically introduced with the help of single-stranded DNA (ssDNA) attached to the particle. Such  DNA-functionalized NPs and colloids are designed to selectively bind to each other, thanks to Watson-Crick hybridization of complementary ssDNA \cite{Gang_DNA,Mirkin_DNA,Francisco}. Even the simplest spherical particles with isotropic DNA-mediated interactions have been shown to form a remarkable range of crystalline lattices \cite{Macfalane,Pine2015}. 

\begin{figure}[ht]
\centering
\includegraphics[width=0.6 \textwidth]{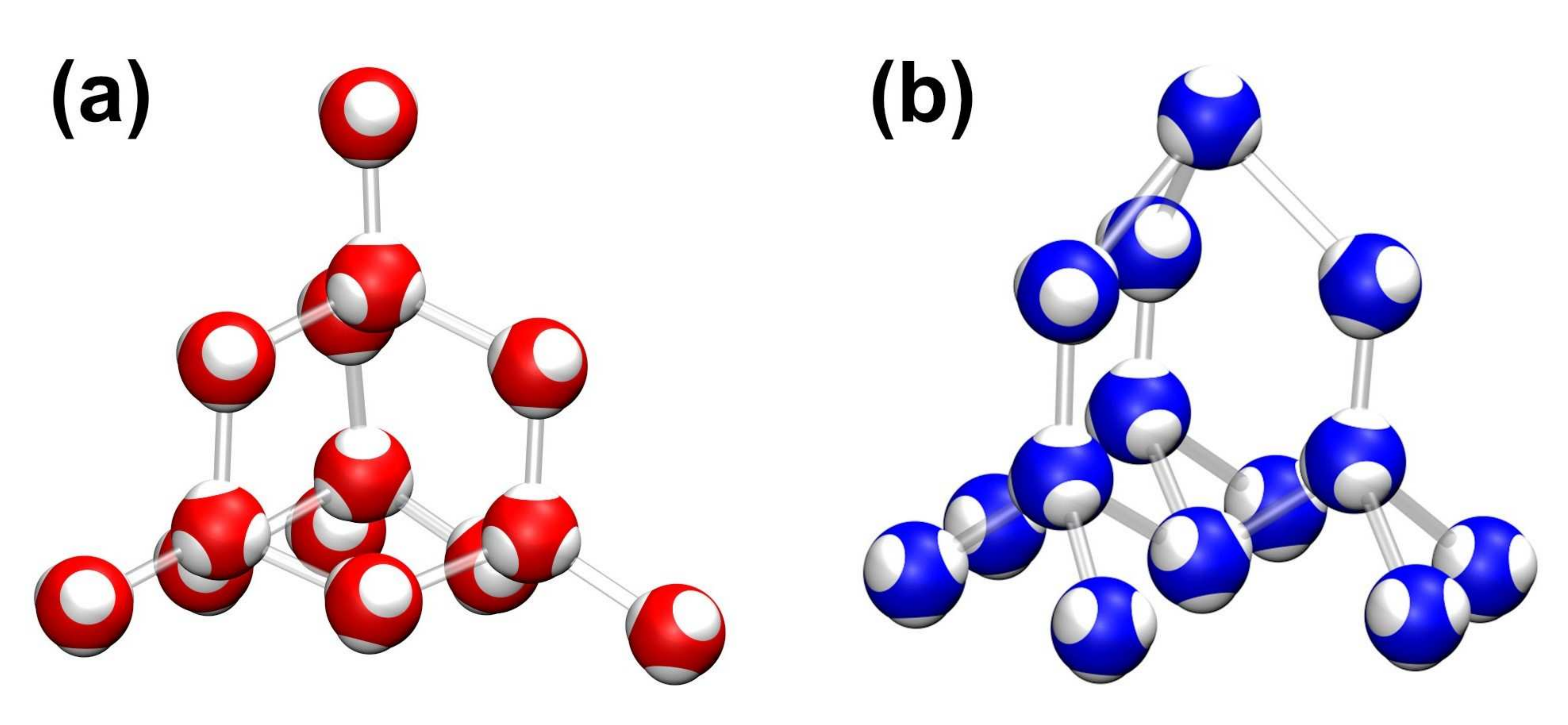}
\caption{Forms of diamond lattice structure: (a) Cubic diamond (CD), (b) Hexagonal diamond (HD).} 
\label{particle1}
\end{figure}

The major reason why the patchy particles are considered to be promising building blocks for self-assembly is that  their local arrangement could be enforced by choosing a specific symmetry of the patch pattern. A classic example would be a diamond lattice, often considered to be  the ``Holy Grail" of colloidal self assembly thanks to its potential for fabrication of photonic bandgap materials \cite{bandgap1,bandgap2}. In a diamond, each particle is supposed to be bound to  four other particles forming  a tetrahedron around it. Diamond is notoriously hard to assemble, partially due to its low density: most self-assembly techniques favor compact arrangements, either to minimize interparticle potential, or to maximize system's entropy. A particle with tetrahedral arrangement of patches is thought to  be a natural building block that would favor local tetrahedral arrangement, consistent with a diamond lattice. However, both Monte Carlo (MC) and Molecular Dynamics (MD) simulations of this system reveal serious fundamental limitations of such an approach \cite{phase_diagram_patchy,Scior_Nat_phys,Glo_diamond,Vasilyev_1,Vasilyev_chrom}. First, there are two forms of the diamond lattices: cubic diamond (CD) and hexagonal diamond (HD), both having local tetrahedral arrangement, as shown in Figure~\ref{particle1}. While CD is more symmetric and slightly more favorable thermodynamically, this competition would lead to  a very slow self-assembly kinetics. Furthermore, CD  lattice has to compete also against a liquid structure with local tetrahedral organization (so-called tetrahedral liquid) \cite{Scior_Nat_phys}. It is predicted that CD becomes a ground state of the patchy particle system only in the limit of sufficiently small patch size, i.e., very strong directionality of the bonds. As a result, the success of using patchy particles for self-assembly of a diamond  or in fact any other non-trivial structure remains limited so far.  

Among the encouraging recent developments is the experimental demonstration of self-assembled CD lattice in the system of DNA-functionalized NPs combined with tetrahedral ``DNA cages" \cite{diamond_oleg}. Those DNA cages are building blocks that are made with the help of DNA origami technique, and can be viewed as nanoscale patchy particles. However, it should be noted that this analogy is not straightforward: the ``patches" at the vertices of the cage  have anisotropic triangular shape which favors global  CD arrangement. This was  consistent with the earlier  numerical results that highlighted an importance of restricting the rotational degree of freedom of the  patch-patch interactions to achieve a robust self-assembly of a CD lattice \cite{Glo_diamond,Sciort_Ncomm2012}.

At present, there are several approaches to experimental implementation of colloidal patchy particles \cite{Pine_patchy_2003,Patchy_fab}. While achieving a non-circular  patch shape remains challenging, the control over their overall arrangement and the particle symmetry is remarkably good. Furthermore, the patches can be decorated with DNA, thus enabling a selective DNA-based interactions between patches \cite{Pine_2012}. Ideally, one would like   to design  so-called ``polychromatic"  particles in which each individual patch is decorated with a specific ssDNA sequence \cite{Vasilyev_chrom}. Our previous studies indicate that this would lead to a high level of control over the resulting self-assembled morphology, and in fact enable the  design of a  nearly arbitrary self-assembled structure \cite{John_PRE2013}. Unfortunately, these hypothetical polychromatic patchy particles are not yet available. Nevertheless, some level of differential DNA functionalization has been achieved. Namely, one can effectively ``color" all the patches of a given particles with one type of DNA, and the rest of the particle surface (the ``shell") with another \cite{Pine_patch_shell}. This already opens a variety of new opportunities. Not only particles can be designed to have different symmetries (e.g. octahedral, tetrahedral, cubic etc.), but one can independently control the strength of patch-patch, patch-shell, and shell-shell attraction between any pair of them.

In this paper, we use Brownian Dynamics (BD) simulations to demonstrate the potential of such particles with differential DNA-functionalized patches and shells, as a platform for  programmable self-assemblies of non-trivial periodic lattices.   Another new element in our approach, compared to the  previous numerical and experimental studies of the patchy particles, is the use of layer-by-layer growth protocol. This is a well known technique in material science and chemical engineering \cite{Caruso-2015,Caruso-2016,Bin-2016}, and as we show below, it can be employed in the context of patchy particles to create periodic structures of exceptional quality.

The specific model system that we consider is motivated by the classic quest for a self-assembled diamond lattice. Similar to other proposals \cite{Glo04,Patchy_cluster,Kern,Scior_Nat_phys,Patchy_fab,Pine_patchy_2003}, we use particles with tetrahedral patch arrangement, consistent with the diamond symmetry, but instead we target a BCC arrangement of four particle types, each forming an FCC sub-lattice. This structure can also be viewed as  Double Diamond (DD) lattices, which enables its conversion to a regular diamond through partial particle deletion. Interestingly, self-assembly of DD lattice has been recently observed in  conventional (non-patch) colloidal system with DNA-mediated interaction, but the mechanism  behind its formation remains obscure \cite{ddiamond_crocker}.

\section{Model and Methodology}

\vspace{-4mm}

\begin{figure}[ht]
\centering
\includegraphics[width=0.6 \textwidth]{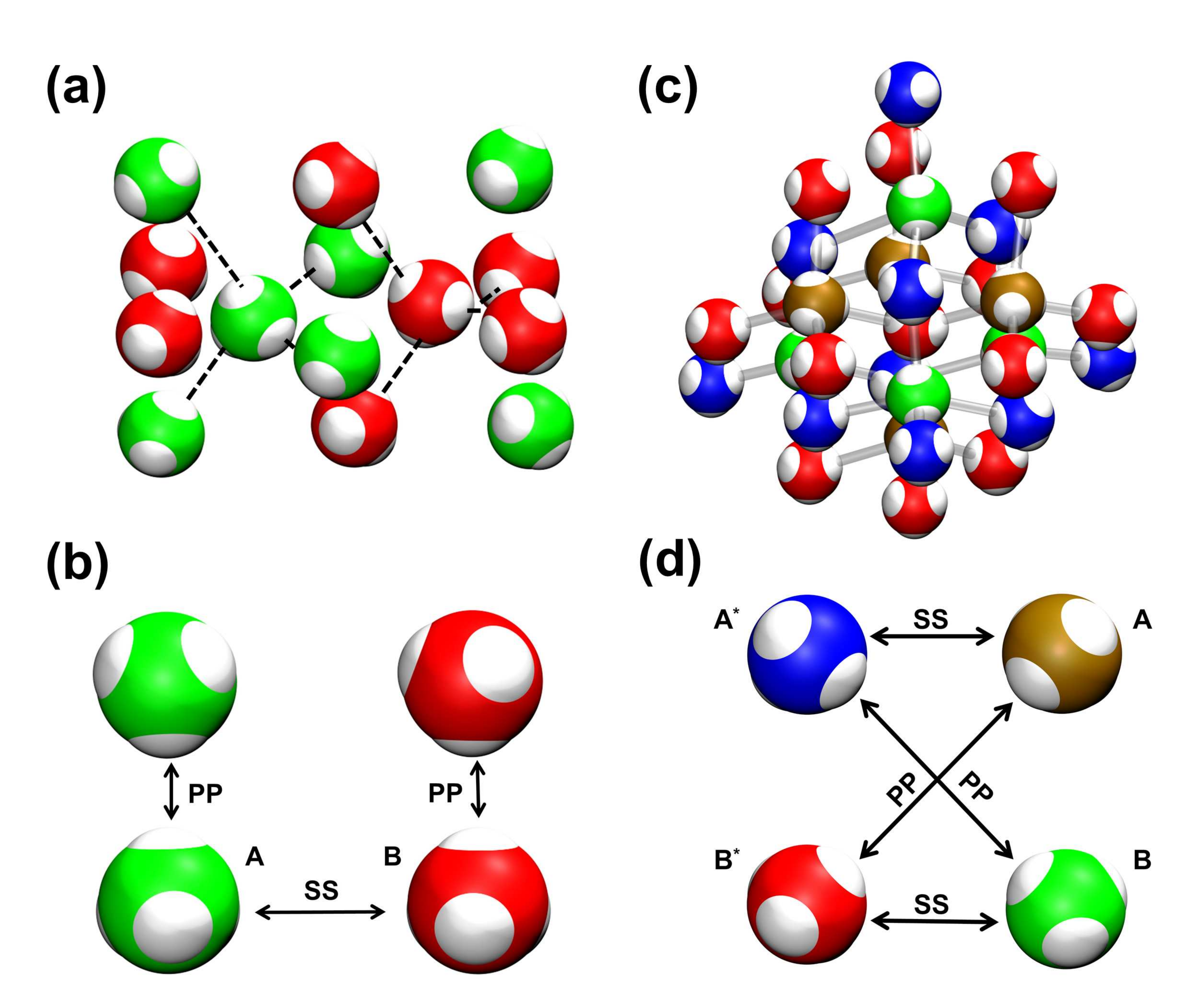}
\caption{ (a) Binary A-B system arranged into Double Diamond (DD) lattice. (b) Interaction rules for assembly of  DD lattice from a binary patchy  system. Patch-patch interaction is introduced between same type of particles (A-A or B-B). Shell-shell interaction is introduced between different types (A-B). (c) DD-derived  lattice for  the quadruple systems. (d) Interaction rules for the quadruple systems.} 
\label{particle2}
\end{figure}

The model structure that we target for self-assembly in this numerical study is a variation of  DD lattice. The latter can be obtained by placing two types of particles: ``A" and ``B" into sites of a BCC lattice in such a way that each particle has four tetrahedrally arranged nearest neighbored  of its type, and four more of the other type, as shown in  Figure \ref{particle2}(a). This structure has a number of advantages over the regular CD  lattice from the point of view of its self-assembly. First, it solves the problem of competition between cubic and hexagonal diamond: only the former allows an interpenetrating double lattice. Second, unlike the diamond itself, this structure  is rather dense thus allowing extra interparticle contacts leading to an additional thermodynamic stability. In fact, this BCC arrangement is known to be a part of the phase diagram of tetrahedral patchy particles, but it requires substantial osmotic pressure to get stabilized \cite{Scior_Nat_phys,phase_diagram_patchy}. 

Rather than relying on an external pressure, one can  propose a scheme in which this structure is favored by interparticle interactions: e.g. patch-patch attraction between particles of the same types (A-A or B-B), and  shell-shell attraction between those of the different types (A-B), as shown in Figure \ref{particle2}(b). By construction, a binary system of patchy particles with tetrahedral arrangement of patches and  these interaction rules, would prefer DD lattice as that  would maximize the number of favorable contacts. However, these particle will also have a tendency to random aggregation, which is a common obstacle to self-assembly of high quality crystalline structures, especially in colloidal and NP systems.    

In this work, in  order to avoid the undesired random aggregates, we propose a layer-by-layer protocol for the self-assembly process. Specifically, we introduce a model system of four particle types: A$^{\star}$, B$^{\star}$, A, and B. The target structure, shown in Figure.~\ref{particle2}(c), is essentially the same as a DD, except each of the diamond sub-lattices is now made of particles of two types. Together, the particles form a BCC lattice made of four FCC sub-lattices, each containing particles of only one type. In order to achieve the self-assembly of this complicated structure,  a pairwise patch-patch interaction is introduced between particle types A$^{\star}$ and B, as well as B$^{\star}$ and A, respectively, as illustrated in Figure ~\ref{particle2}(d). A pairwise shell-shell interaction is introduced between particle types A$^{\star}$ and A, as well as B$^{\star}$ and B. The advantage of this interaction scheme is that now the particles can be introduced in batches: A$^{\star}$+B$^{\star}$ or A+B, and there is no attractive interactions between the particles of the same batch (there is a regular hard core repulsion between any two particles). This opens a possibility of a very clean layer-by-layer assembly, not affected by aggregation of the free particles in the suspension. 

\begin{figure}[ht]
\centering
\includegraphics[width=0.8 \textwidth]{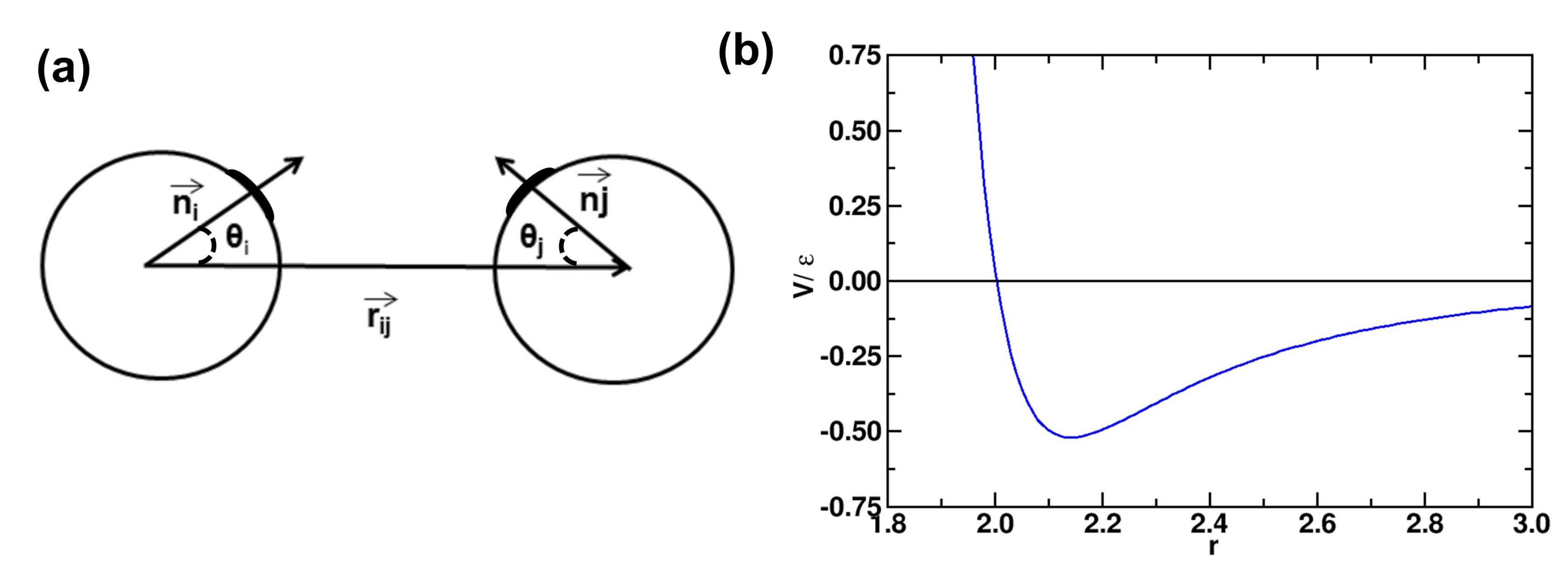}
\caption{(a) Interacting patchy paricles.($\vec r_{ij}$) is the vector connecting  their centers. Angler $\theta_{i}$ and $\theta _{j}$) determine the patch orientation with respect to the centerline. (b) Interparticle patch-patch potential vs center-to-center distance  $r$, for the case when the patches are facing each other. } 

\label{particle3}
\end{figure}

To study the system, we introduce the force  field which is analogous to Kern-Frenkel \cite{Kern} model of patchy interactions,  but replaces step-like potentials with smooth functions making it  suitable for Molecular Dynamics (MD) or BD  simulations. A number of conceptually similar but formally distinct  force fields have been previously proposed  in Refs.  \cite{Caccuiuto-2009,Panagiotopoulos-2016}. 
The patch-patch (PP) and shell-shell (SS) pairwise interaction potentials are modeled as:
\begin{eqnarray}
 V_{PP} = - \epsilon_{PP} \left(\frac{2a}{r}\right)^{6} \ g(\theta_{i}) \ g (\theta_{j})
\label{eqnPP}
\end{eqnarray}
\begin{eqnarray}
V_{SS} = - \epsilon_{SS} \left(\frac{2a}{r}\right)^{6} \ (1-g(\theta_{i})) \ (1 - g (\theta_{j})),  
\label{eqnSS}
\end{eqnarray}

Here $\epsilon_{PP}$ and $\epsilon_{SS}$, are patch-patch and shell-shell interaction  parameters, respectively. $g(\theta) = \left(\frac{1}{1+e^{\alpha(p_{0}-p)}}\right)$ is the angular potential and  $\alpha$ is its  sharpness parameter. $g(\theta) \approx 1$ when particles i and j are oriented so that the vector joining their centers ($\vec r_{ij}$) intersects an attractive patches on both particles; otherwise $g(\theta) \approx 0$. $p= cos\theta= \vec n_{i} \ . \ \hat{r}_{ij}$, where $\theta$ is an  angle between  vector  $\hat {r}_{ij}=\vec r_{ij}/|\vec r_{ij}| $,  and the unit vector  ($\vec n_{i}$) that points to  the center of a patch  from  the center of  particle $i$, as shown in Figure~\ref{particle3}a. $p_{0} = cos\theta_{0}$ is a constant  that determines the patch size. The fundamental length scale in our problem is the  particle radius, a = 1.00. The unit time is the self-diffusion time $\tau = \frac {a^2}{6D}$, where $D$ is the diffusion coefficient of the particle \cite{John_PRE2013}. All energies are given in units of $k_BT$. 

In addition to attraction, there is a type-independent hard core repulsive potential:
\begin {eqnarray}
V_{Rep}(r)= \epsilon \ e^{-\beta(\frac{r}{a}-2)}.
\label{eqnRep}
\end{eqnarray}
Here $\epsilon=(\epsilon_{PP}+\epsilon_{SS})/2$. In our simulations we keep the relation between $\beta$ and $\alpha$ as $\alpha = 4 \  \beta \ \frac{1}{\sqrt{(1-(p_{0}^{2}))}}$, to ensure that the maximum force generated by the attractive and repulsive potentials are comparable. Specifically, we use $\beta = 15.00$ and $\alpha = 124.04$. The  particles undergo translational and rotational diffusion subjected to the forces and torques derived from the above interaction potentials, Eqs. \ref{eqnPP}-\ref{eqnRep}.

\section {Layer-by-Layer assembly}
\begin{figure}[ht]
\centering
\includegraphics[width=0.6 \textwidth]{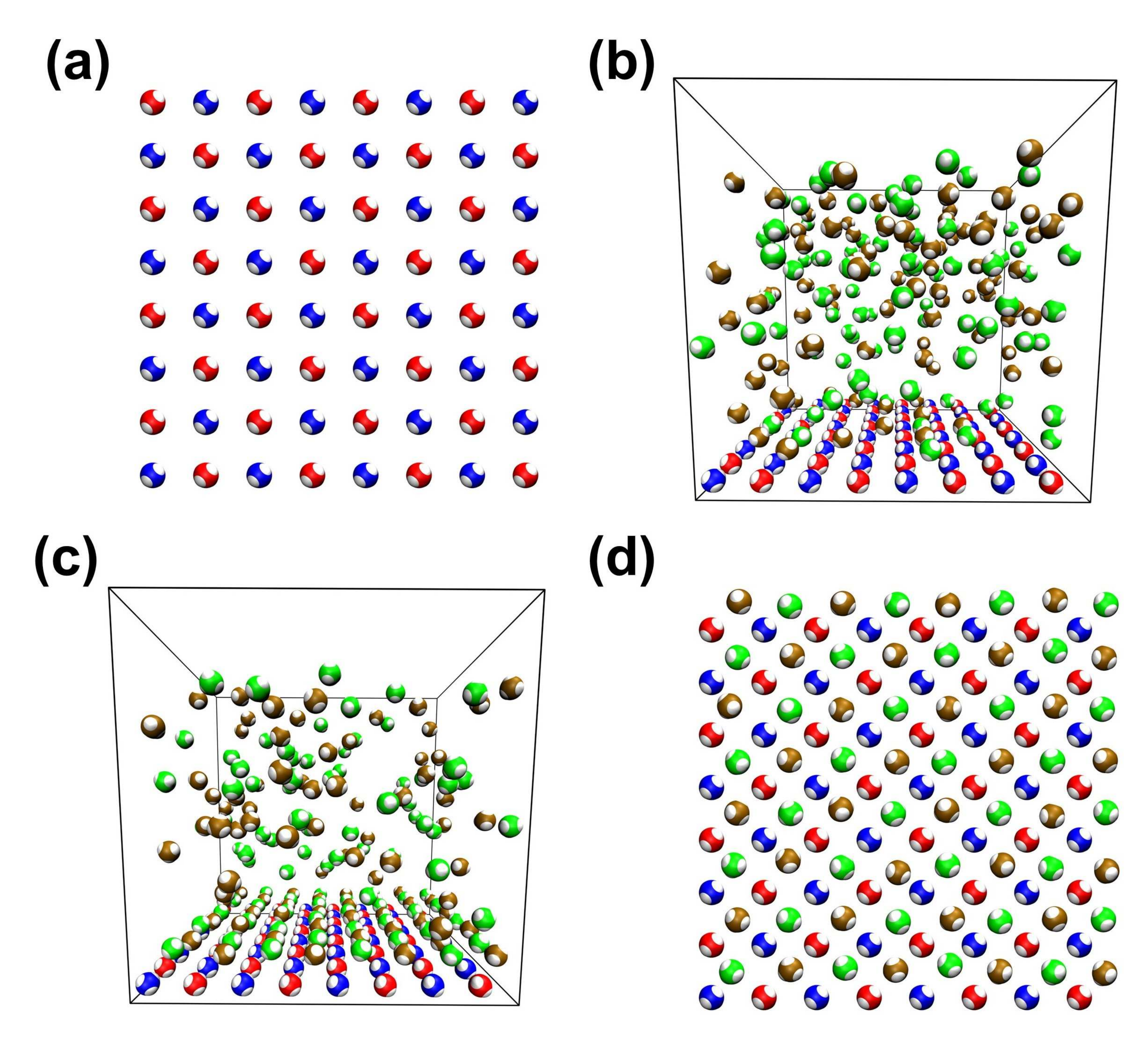}
\caption{Formation of layer 1 on top of the template layer. (a) A periodic square lattice, made of A$^{\star}$ and B$^{\star}$  particles, is used as a template. (b) Free A and B  particles are introduced. (c) At time $t= 1000$, the new  layer is  fully formed. (d) Top view of the adsorbed A and B particles on top of the template layer. Free particles are discarded.}
\label{lay1}
\end{figure}

We now  introduce the layer-by-layer assembly protocol. This approach is  extensively used  in various sub-fields of  material science and chemical engineering \cite{Caruso-2015,Caruso-2016,Bin-2016}. A periodic square lattice consisting  of particles  A$^{\star}$ and B$^{\star}$, acts  as a template,  as shown in Figure~\ref{lay1}(a). Two patches of each particle are oriented upward,  diagonally to the square unit lattice. Particles of the template layer are fixed during the  simulation. The radius of the particle, $a=1.00$ and the lattice constant is $2.50a$ (the value is chosen so that a particle  would fit  well in a BCC unit cell).  Next, we place $n=180$ free  A- and B-type particles (number ratio 1:1) inside  the cubic simulation box that already contains the template layer. The box side  is $L=20.0$, and all the particles are subjected to wall-like boundary constraint that effectively limits the box height to $h=0.9L$. The  volume fraction of the free particles is $\delta = \frac {4}{3} \pi  a^{3} n \frac {1}{L^2h}$ = $0.10$.  The interactions between the particles are given by the rules shown in  Figure~\ref{particle2}(d), and interaction potentials  Eqs. \ref{eqnPP}-\ref{eqnRep} with the following parameters:  $\theta_0 =40 ^\circ$, $\epsilon= \epsilon_{PP}=\epsilon_{SS} = 8.00$. 
The wall at position $z=h=0.9L$ is introduced in the form of a local downward  force  $f_z = -(\frac{a}{z - h})^{12}$. 

At time $t = 1000 $, the system  reaches a steady state with a new layer of A and B particles formed on top of the template, as shown in Figure~\ref{lay1}(c). One can see that e.g.  B particles  (green color) binds  with   A$^{\star}$ particles (blue color) through PP interaction and with B$^{\star}$ (red color) through SS interaction, as dictated  by our design. Figure~\ref{lay1}(d), shows the  the top view of  A and B  particles adsorbed on top of the template layer. 

At the next step, the  adsorbed particles are kept  while  all other A and B particles are discarded, and the new batch of  particles of types A$^{\star}$ and B$^{\star}$ is introduced. To maintain the same  volume fraction,  the position of the upper boundary $h$ is shifted upward by the amount $1.3a$, which is determined as a vertical separation between the first and second layer. Similarly to  the previous step, the BD simulation is run for time $t=1000$ sufficient for the particles to form the new layer.  After that,  all free particles are again discarded, and the new batch  is introduced, this time of types A and B again. This procedure can be repeated indefinitely, each time resulting in a formation of the new layer. To speed up the simulation, we ``freeze" positions and orientation of  all the particles that are  located deeper than two layers from the surface.

\begin{figure}[ht]
\centering
\includegraphics[width=0.6 \textwidth]{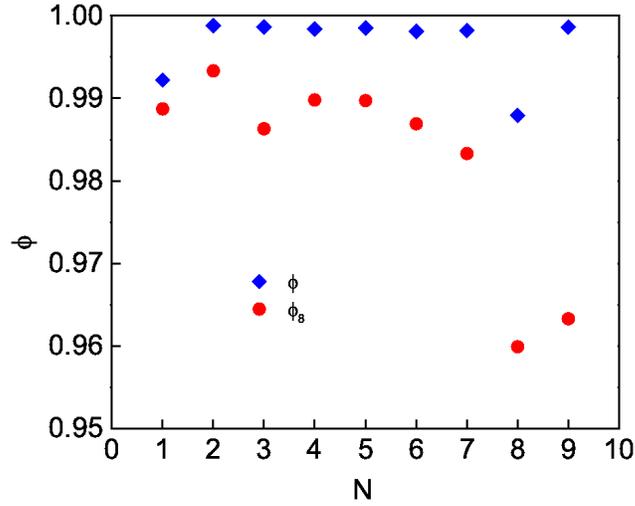}
\caption{Filling factors $\phi$ (blue diamonds) and $\phi_8$ (red circles), for each of the deposited layers, $0<N<10$.}
\label{lc1}
\end{figure}
\begin{figure}[ht]
\centering\includegraphics[width=0.6 \textwidth]{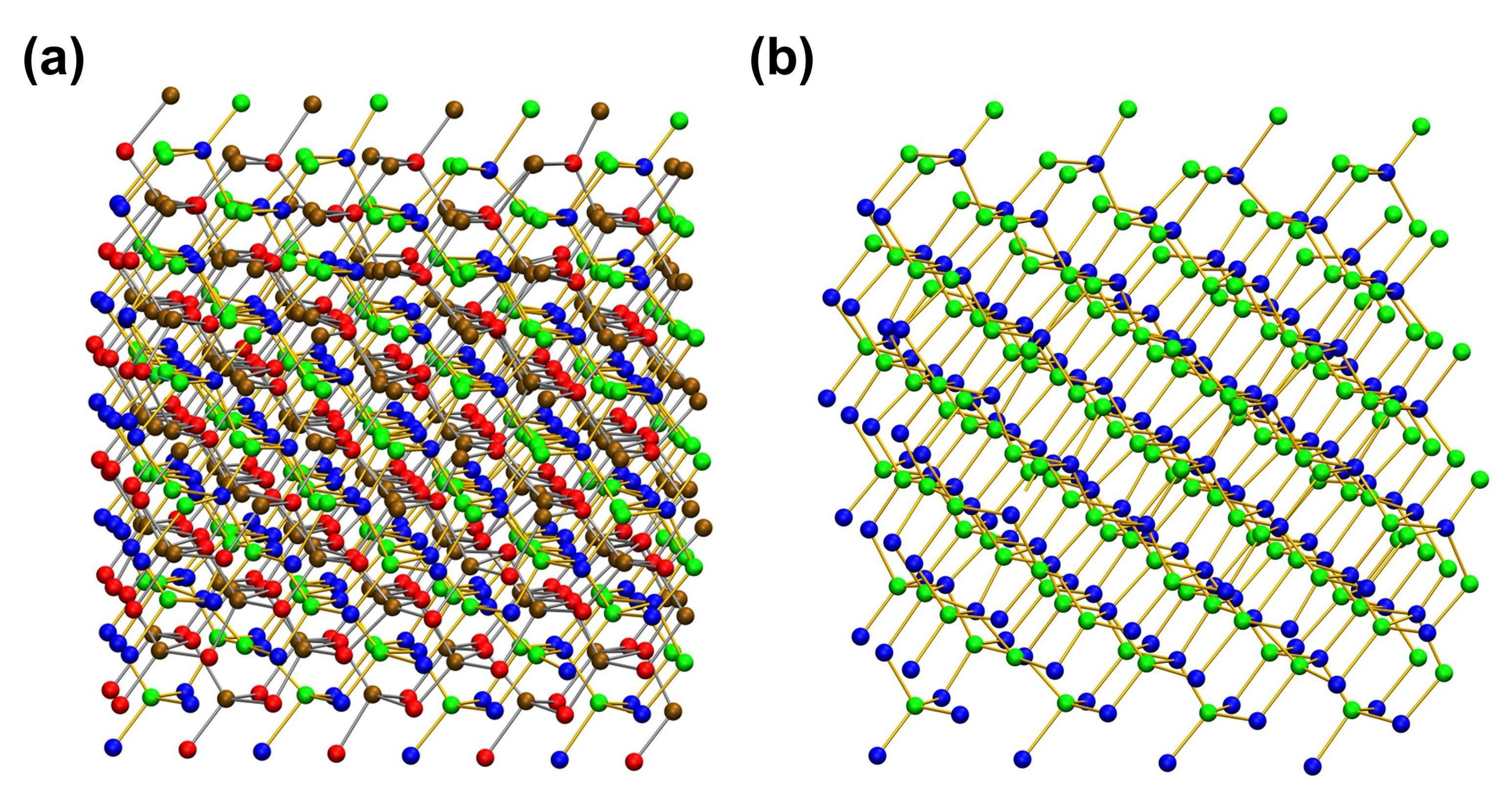}
\caption{Formation of DD lattice by layer-by-layer assembly growth process. a) Two interpenetrating diamond lattices form DD lattice (patches are remove for better view). b) Diamond lattice is obtained after removing one of the interpenetrating diamond lattices (patches are removed for better view).}
\label{dd1}
\end{figure}

In order to check the quality of the obtained structure, we calculate several quantities that work as local order parameters. First, we can determine the filling factor  $\phi$  in a particular layer, which is defined as the ratio of the number of particles in that layer  to the number of sites that they can fill.  Second, we can determine  coordination number $Z$ of each particle by counting the contacts with its  neighbors, but only those  which are consistent with the target structure. This allows one to calculate anothe filling factor,  $\phi_{8}$,  by counting  the particles in   each layer that have the maximum number of such contacts, $Z=8$. Naturally, this parameter is only meaningful  for the inner layers but not for the top  one. The results are averaged over ten independent runs.  As shown in Figure~\ref{lc1}, the overall filling factor  $\phi$ is as high as  $0.99$ for all the layers,  whereas parameter $\phi_{8} \geq 0.98$ is also quite close to 1.  

By using the layer-by-layer protocol, we self-assembled the patchy particles into a ten-layer structure as shown in Figure~\ref{dd1}(a) (the patches are removed for better view). As intended, the simulations produced a high quality lattice that can be interpreted as DD structure made of two  interpenetrating CD sub-lattices. After removing one of them formed by the  particles of types B$^{\star}$ and A, we obtain the CD lattice shown in Figure~\ref{dd1}(b).  

\section{Robustness of Layer-by-Layer Approach}
\begin{figure}[ht]
\centering
\includegraphics[width=0.7 \textwidth]{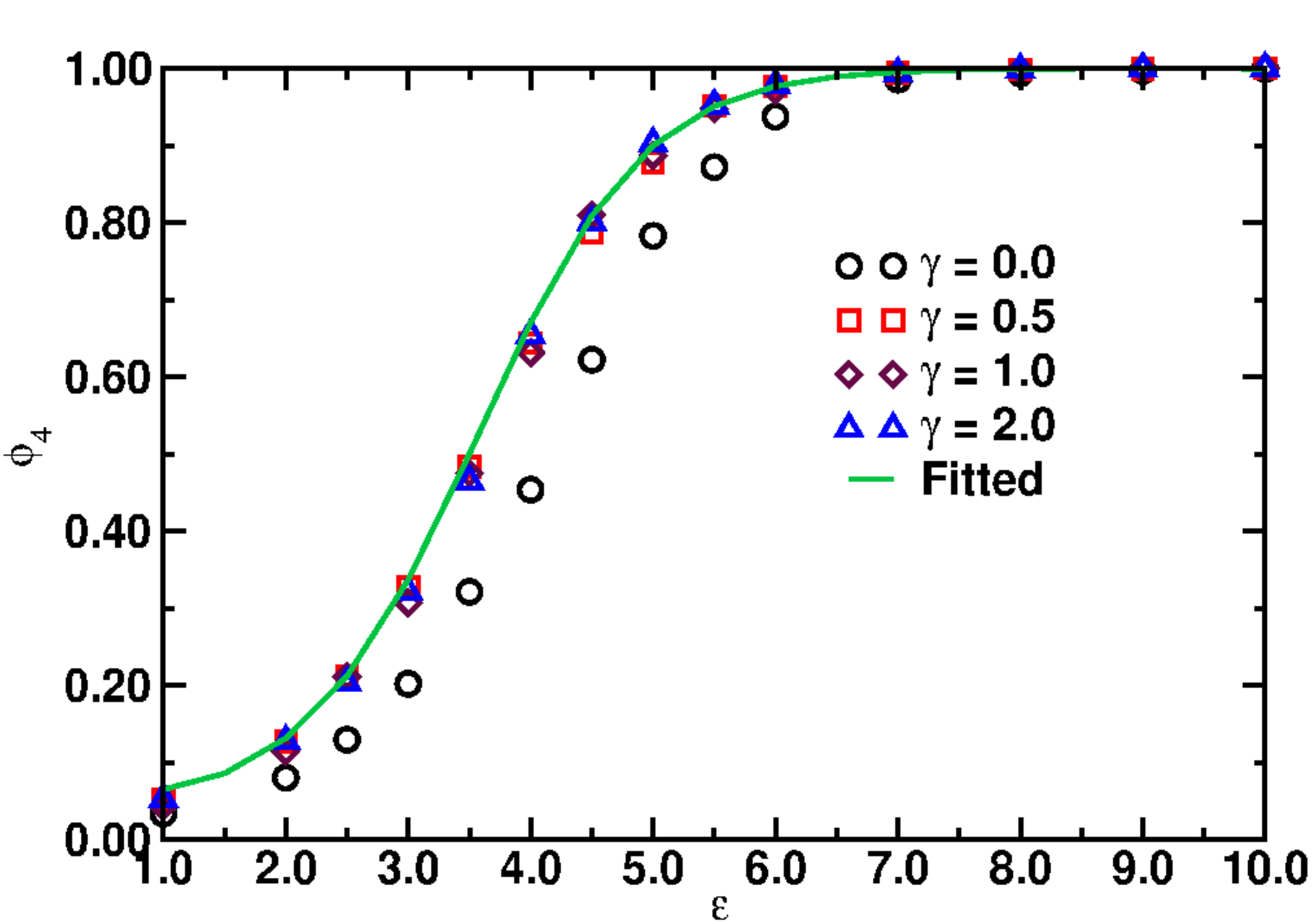}
\caption{Filling factor $\phi_{4}$ of layer 1 for different values of  $\gamma$ and constant patch size,  $\theta_{0}$ ($=40^\circ$). The solid curve is given by Boltzmann-Gibbs distribution,  Eq.~\ref{Theo2}.} 

\label{Angle40}
\end{figure}
In order to verify the robustness of our proposed strategy, we first investigate the adsorption of the  particles on top of the template layer across a wide range of the model parameters. Namely, we choose different values of $\theta_0$:  $10^\circ$, $25^\circ$, $40^\circ$, and $50^\circ$. We also vary the interaction  strength of PP and SS interaction. To do this,  we  set  their ratio  $\gamma=\frac{\epsilon_{SS}}{\epsilon_{PP}}$  at values  $0.0, 0.5, 1.0$, or  $2.0$, and  gradually vary the average interation strength,  $\epsilon=\frac{\epsilon_{SS}+\epsilon_{PP}}{2}$.  All other conditions are kept the same as in the original simulations. 

Figure~\ref{Angle40} shows the filling factor $\phi_{4}$ for the first adsorbed layer made of A and B particles as a function of average binding strength $\epsilon$, for various values of  $\gamma$, and fixed patch angle $\theta_0=40^\circ$. Similarly to $\phi_{8}$, $\phi_{4}$ accounts for the  particles that have the maximum number of the favorable contacts, which in this case is $Z=4$.   As one can see from the Figure, almost all the data  collapse onto a single curve, with an exception of the case of $\gamma=0$ (no shell-shell interactions) which is slightly shifted. The crossover from the empty to completely  filled layer occurs around $\epsilon=4$, and the  curves saturate for  $\epsilon>7$.    This result can be  described theoretically with a simple two-state model, in which  each site can be either empty or occupied, and the free energy of the  ``occupied" state is dominated by the strength of interaction of the particle  with its  Z neighbors, $Zk\epsilon$. Here $k\epsilon=k(\epsilon_s+\epsilon_p)/2$ is the average of the depths of  SS and PP potentials, with  constant  $k=0.520$ determined  from  Figure~\ref{particle3}(b). We can successfully fit the simulation data with a regular Gibbs-Boltzmann formula: 

\begin{eqnarray}
%
\phi_{4}=  \frac{1}{1+(\frac{\epsilon}{\epsilon_0})^{\frac{3}{2}} e^{-Zk(\epsilon - \epsilon_0)} },
\label{Theo2}
\end{eqnarray}

\begin{figure}[ht]
\centering
\includegraphics[width=0.7 \textwidth]{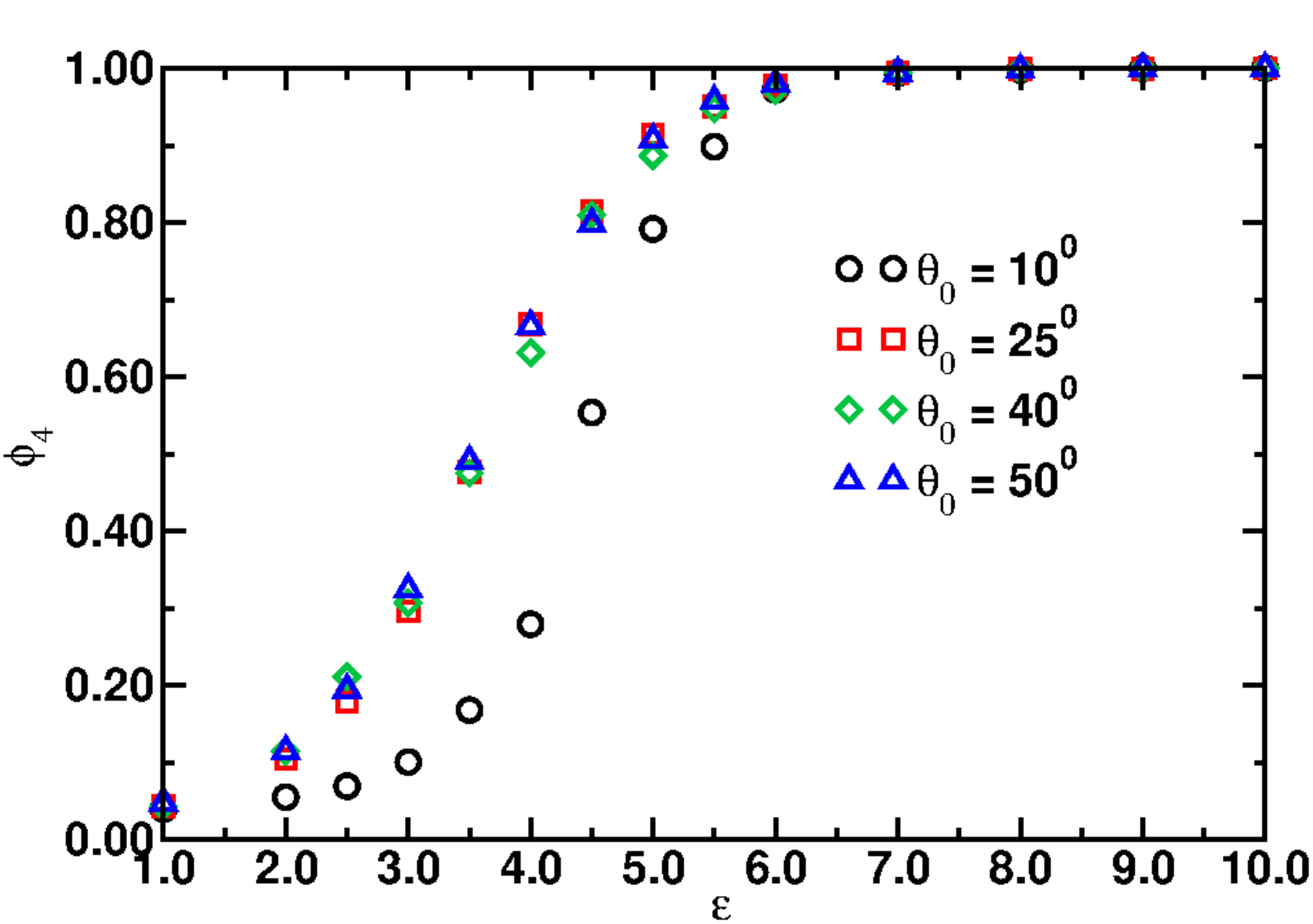}
\caption{Filling factor $\phi_{4}$ of layer 1 for different vales of  $\theta_{0}$ and constant  $\gamma$ ($=1.0$).} 

\label{AngleG1}
\end{figure}

The best fit is achieved for  $\epsilon_{0} = 3.5$   and  $Z = 3.5$, which is close to the  expected value of $Z=4$. Note that the prefactor $(\frac{\epsilon}{\epsilon_0})^{\frac{3}{2}}$ accounts for  the fact that   there are three translational degrees of freedom that are localized due to  binding, and the  rigidity of the bonds scale as $\epsilon$.   Similarly, we  investigated the first  layer formation for various $\theta_0$ while keeping $\gamma =1.0$ constant, as shown in Figure~\ref{AngleG1}. Once again, most of the data collapse to the same curve, with an exception of  the smallest patch size, $\theta_0 =10^\circ$.

\begin{figure}[ht]
\centering
\includegraphics[width=0.8\textwidth]{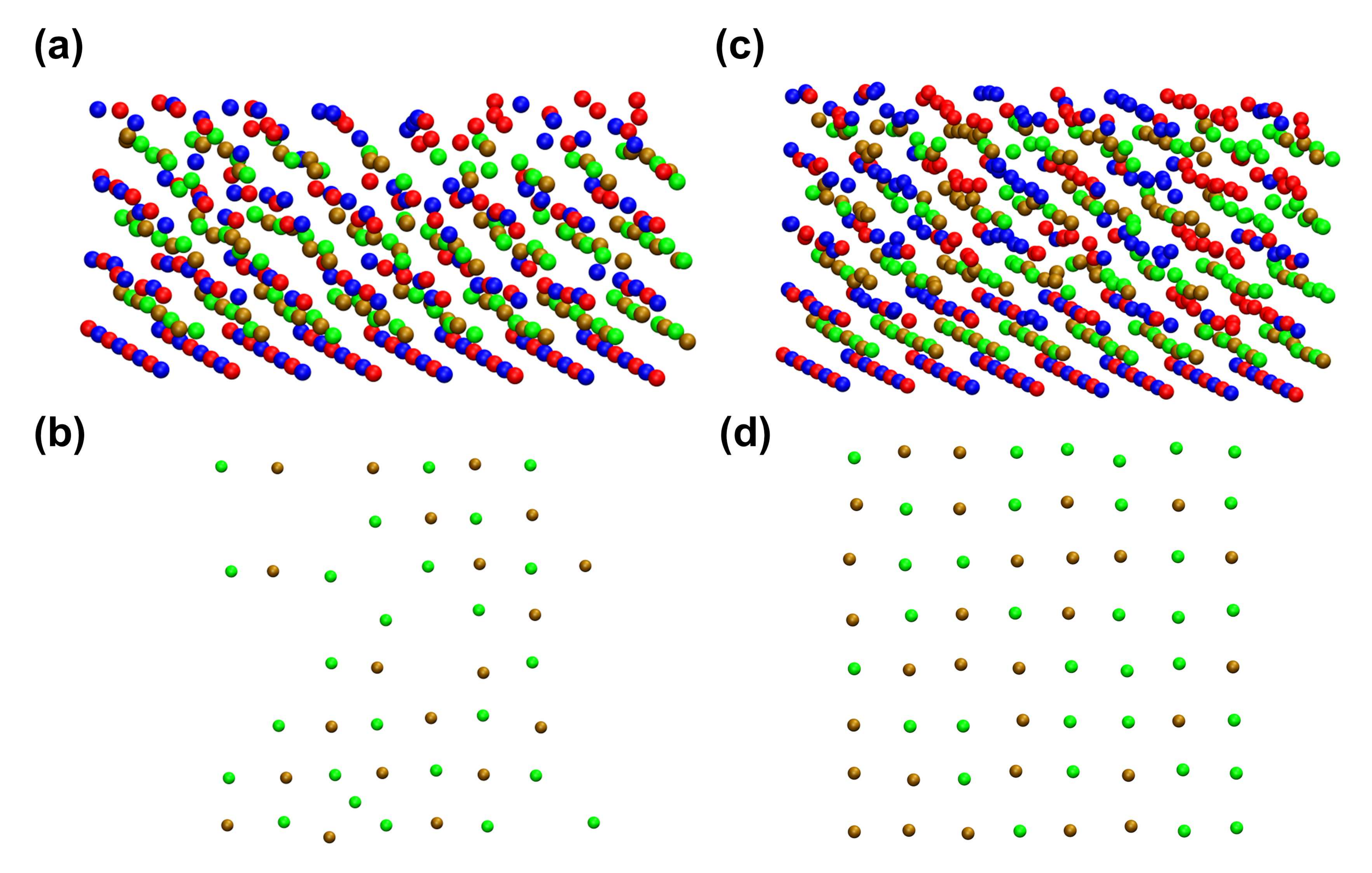}
%
\caption{Formation of DD lattice by layer-by-layer assembly growth process at different $\epsilon$ ($6$ and $20$). a) Two interpenetrating diamond lattices form DD lattice at $\epsilon = 6$ (patches are remove for better view). (b) Not all sites  are  occupied in the layer for  $\epsilon = 6$. (c) Result of self-assembly  at $\epsilon = 20$. (d) Sites are occupied by ``wrong" types of particles for   $\epsilon = 20$ . (Patches are removed for better view)}
\label{E6-20}
\end{figure}
\begin{figure}[ht]
\centering
\includegraphics[width=0.5\textwidth]{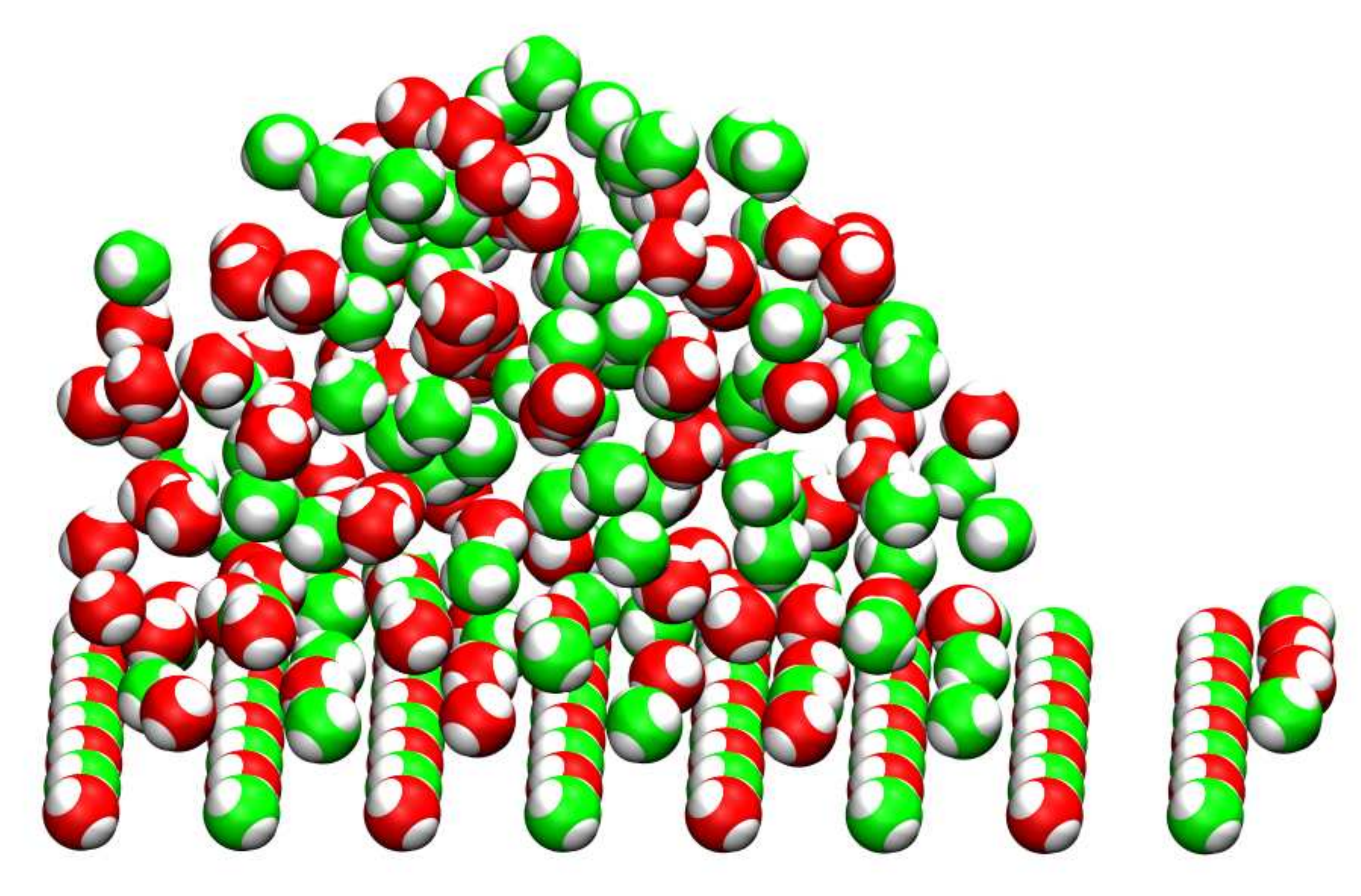}
%
\caption{Formation of random aggregate  in   AB system. }
\label{AB}
\end{figure}

We found that once the first layer is fully saturated (for $\epsilon>7$), the complete  layer-by-layer procedure yields a near perfect lattice  in a relatively  wide  window of parameter $\epsilon$, at least up to $\epsilon = 10$. However, once the interaction strength is very large, the binding becomes nearly irreversible on the time scale of the simulations. This leads to the generation of errors.  Figure~\ref{E6-20}(a-d) illustrate how the procedure  fails outside the  preferred parameter window. On the one hand, for $\epsilon = 6$, each layer is not completely filled resulting in a large number of vacancies. On the other hand, in the limit of  a very large interaction parameter, $\epsilon = 20$, the structural BCC order is preserved. However,  the structure becomes compositionally disordered, i.e.  all the sites are occupied but not by the particles of the  right type   (see Figure~\ref{E6-20}(d)). 

\begin{figure}[ht]
\centering
\includegraphics[width=0.5\textwidth]{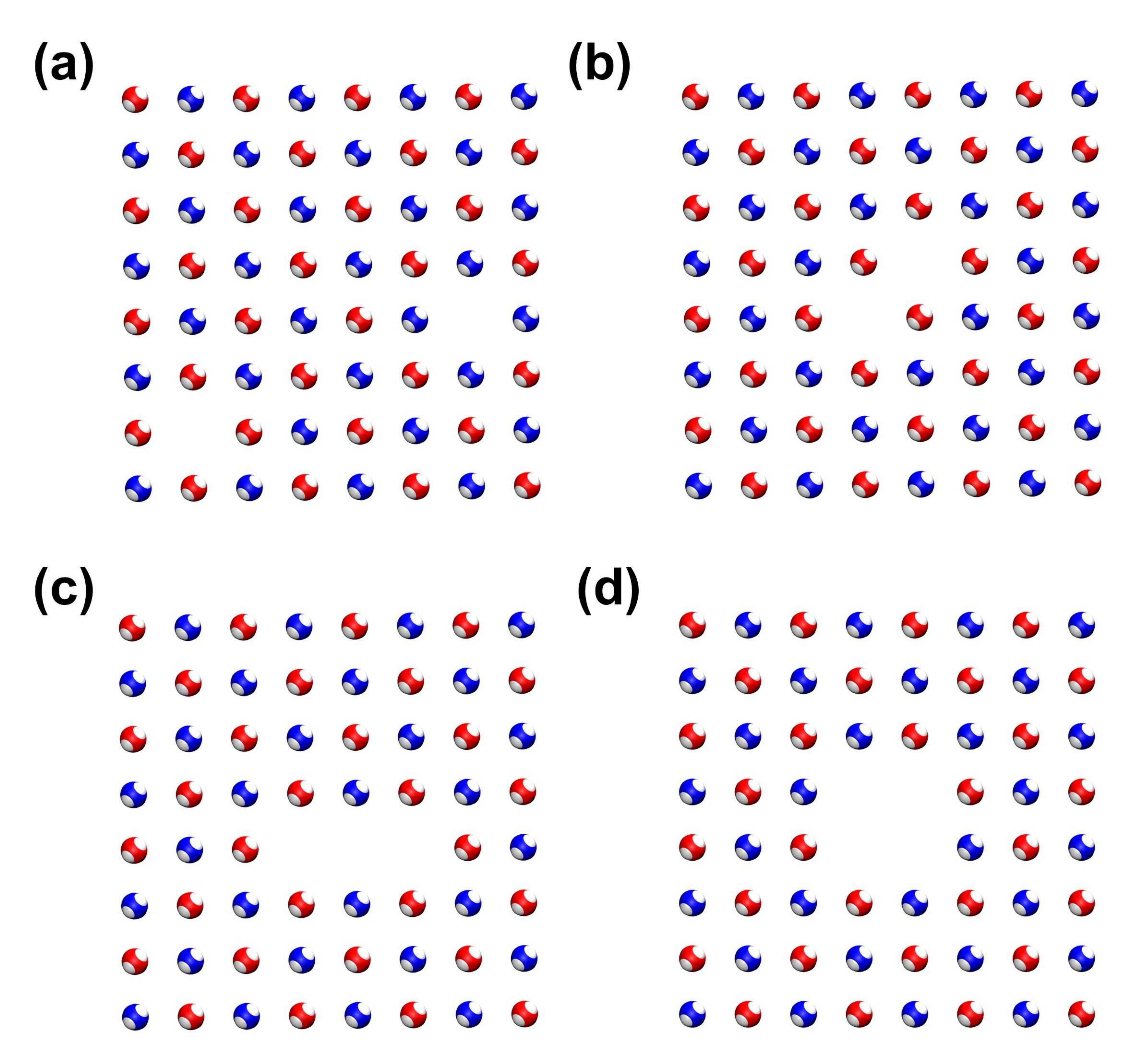}
\caption{Template layers with different types of defects. (a) Two non-adjacent particles are removed from the template layer. (b) Two adjacent particles are removed from the template layer. (c) Three adjacent particles are removed from the template layer. (d) Four particles  are removed from the template layer.}
\label{def1}
\end{figure}

When proposing the layer-by-layer protocol and the 4-particle scheme, we argued that the simpler  binary system has a generic  tendency to form random aggregates during the self-assembly process. In order to demonstrate this explicitly,  we simulate a  system made of A and B  particles with interaction rules shown in Figure \ref{particle2}(b). The template layer is also made of A and B  particles. We add free  particles by keeping  volume fraction and other parameters the same as  in the original simulations. As shown in Figure~\ref{AB}, the random aggregate indeed forms  on top of the template layer. By varying the interaction strength $\epsilon$ one cannot achieve the regime in which the ordered structure is growing while the  random aggregate does not.   In principle, by choosing volume fraction substantially lower than in our simulations, one could expect that the ordered structure  nucleated by the template layer,  would require a lower value of $\epsilon$ to be formed, than the one needed for aggregation. However, that scenario appears to be  unrealistic since   the  that volume fraction is  too low and  the kinetics is many orders of magnitude slower than in our scheme.

\begin{figure}[ht]
\centering
\includegraphics[width=0.8\textwidth]{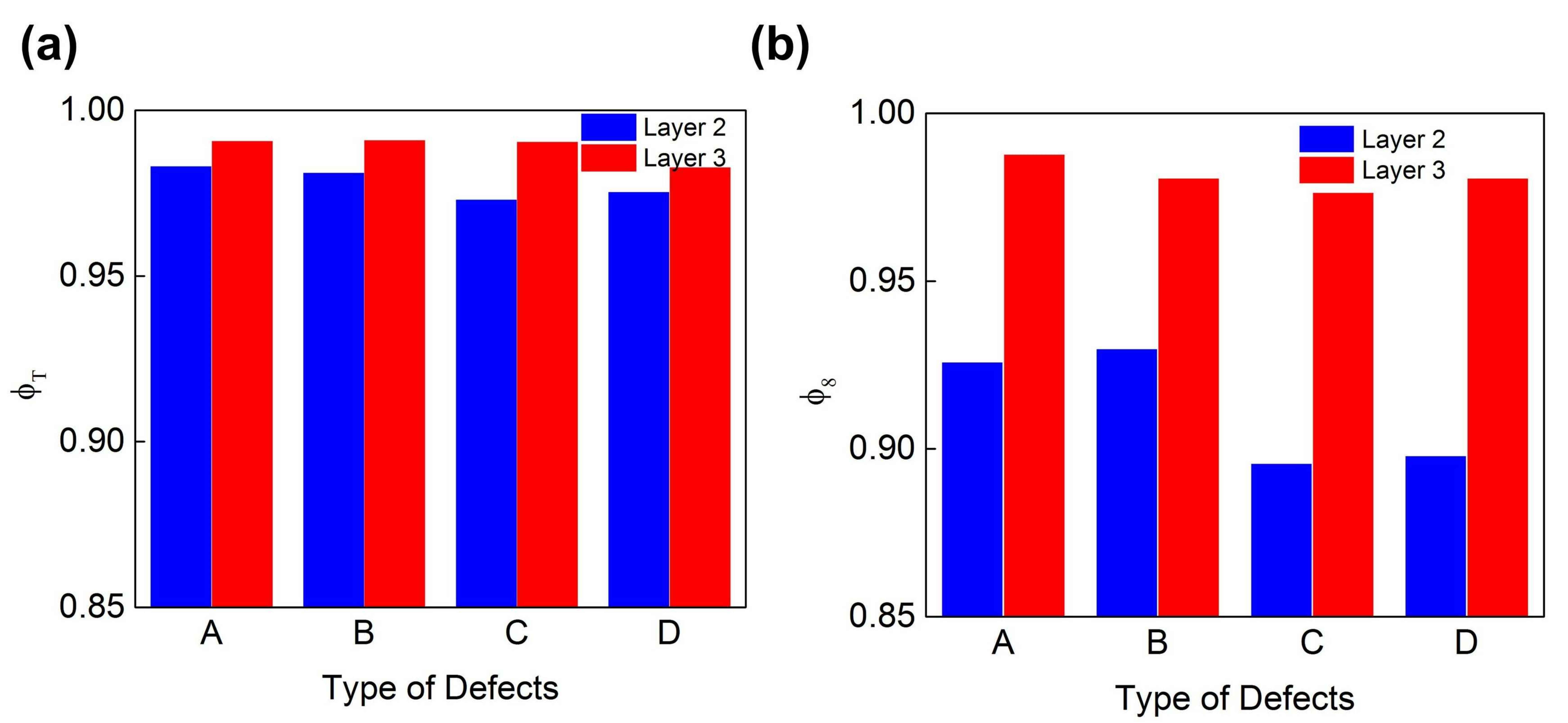}
%
\caption{Layer filling factor ($\phi$) for template layers with different types of defects. (a) $\phi$ for layer 2 and layer 3. (b) $\phi_{8}$  for layers 2 and 3.}
\label{def-ana}
\end{figure}

Finally, we investigate the robustness of the proposed self-assembly strategy by introducing various defects in the template layer, shown in Figure~\ref{def1}. We grow the layers using the above  layer-by-layer protocol while keeping all the  simulation conditions the  same as before, and  check the quality of the obtained layers by determining $\phi$ and $\phi_{8}$. As one can see in Figure~\ref{def-ana}(a), the average  filling factor $\phi$ for the adsorbed layers reaches  $0.98$ after just three layers are deposited,  for all the defect arrangements. Similarly, $\phi_{8}$ is recovered,  as  shown in Figure~\ref{def-ana}(b). The obtained results  show that  such  errors can be ``healed" after depositing of just two to three layers.

\section{Conclusions}

In this paper, we proposed and numerically investigated several strategies that allow to achieve a self-assembly of non-trivial high-quality structures. Specific model example was a BCC lattice made of four particle types, which  can be viewed as a variation of DD lattices. These strategies include: (1) use of selective patch-patch and shell-shell interactions and (2) layer-by-layer assembly protocol. Thanks to a combination of the two approaches, at each stage the growing structure has an exposure to the particles that interact with its reactive surface, but not with  each other. As a result, this type of  self-assembly has a very low error rate. In fact,  we have demonstrated that as the assembly progresses, the occasional errors get corrected and defects ``healed". The proposed scheme is certainly not limited to the specific morphology that we have explored, but this particular structure is of an additional interest since it can be converted to cubic diamond lattices upon selective deletion of half of the particles. 

Not only does the layer-by-layer approach help to  avoid formation of  unwanted phases, such as a random aggregate, it provides a route to self-assembly of structures that {\it do not} have to be a true equilibrium state. For example, we successfully assembled DD lattice even in the regime when shell-shell interactions were completely switched off ($\gamma=0$). In that case, the attractive patch-patch interactions are acting only within pairs A-B$^\star$ and B-A$^\star$. Since the patch size is rather large, the expected equilibrium phase in both of these sub-systems would in fact be a tetrahedral liquid \cite{Scior_Nat_phys} rather than the obtained DD-type lattice. One can imagine a variation of our approach that leads to programmable assembly of aperiodic ``sandwich" structures made of even larger variety of particle types. In fact, {\it it provides a hybrid strategy that combines the  top-down fabrication with bottom-up assembly}.


From the point of view of practical realization of this layer-by-layer assembly strategy, one has to solve two important experimental challenges. First, one has to develop an initiation procedure that creates the first layer for the future layer-by-layer growth. One can imagine  a number of plausible approaches, e.g.  use of DNA microarrays or 2D micro-patterning for positioning of those particles, or a combination of these techniques. Another technical challenge is the automation of the growth procedure: each layer is build by exposing the structure to a specific mixture of particles (either A+B or A$^\star$ + B$^\star$ in our case). Therefore, the environment has to be time-modulated.  This can be achieved either by physically moving the substrate from one environment to another, or by using microfluidics. Development of such automatic protocols will also open new opportunities of  manufacturing both periodic and  aperiodic layered structures with new functionalities, e.g. for photonic applications.

{\bf \MakeUppercase{Acknowledgment}} 

The research was carried out at  the Center for Functional Nanomaterials, which is a U.S. DOE Office of Science Facility, at Brookhaven National Laboratory under Contract No. DE-SC0012704.

{\bf \MakeUppercase{ Appendix: Verification and characterization of the model}}

\begin{figure}[ht]
\includegraphics[width=0.6 \textwidth]{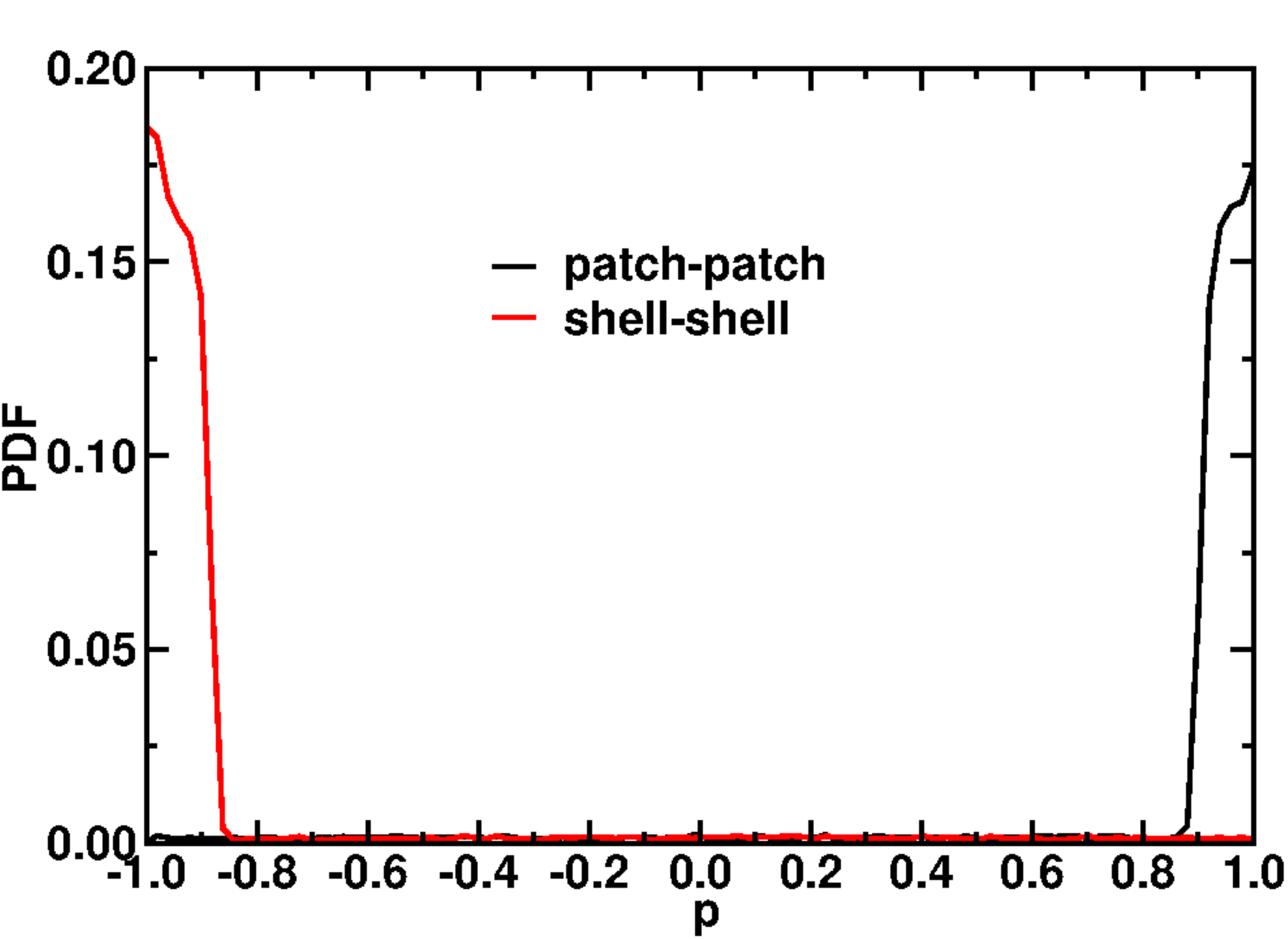}
\caption{Probability distribution function (PDF) of dimer formation. In case of patch-patch interaction (black line), dimer formations only occur when $p (=cos\theta) \geq 8.866$ and while for shell-shell interaction, dimer formations observe at $p \leq 0.866$ (red line).} 
\label{particle4}
\end{figure}
In order to verify that the force field derived from the potentials Eqs.~\ref{eqnPP}-\ref{eqnRep} are working properly, we investigate the dimer formations in a binary systems of mutually attractive single-patch particles. First, we investigate the dimer formation when there is only patch-patch attraction between the particles of different types. Here, we choose  $\theta_0=30 {^\circ}$ ($p_0=\cos\theta_0=0.866$), $\epsilon=15$,  an  monomer volume fraction ($\delta=0.02$. We find that particles only form dimers when the $\theta \leq 30 ^{\circ}$. By analyzing the resulting histogram of $p_{i} =\cos\theta$, we conclude that the orientations of the particles is in the expected range for the PP interaction, $p_{i} \geq 0.866$ (see Figure~\ref{particle4} (black line)). As expected, we observe a very similar histogram, invert with respect to $p_{i}=0$  when only SS attraction is present, with $\theta_0=150 {^\circ}$ (Figure~\ref{particle4} (red line)) .  

\begin{figure}[ht]
\centering
\includegraphics[width=0.6 \textwidth]{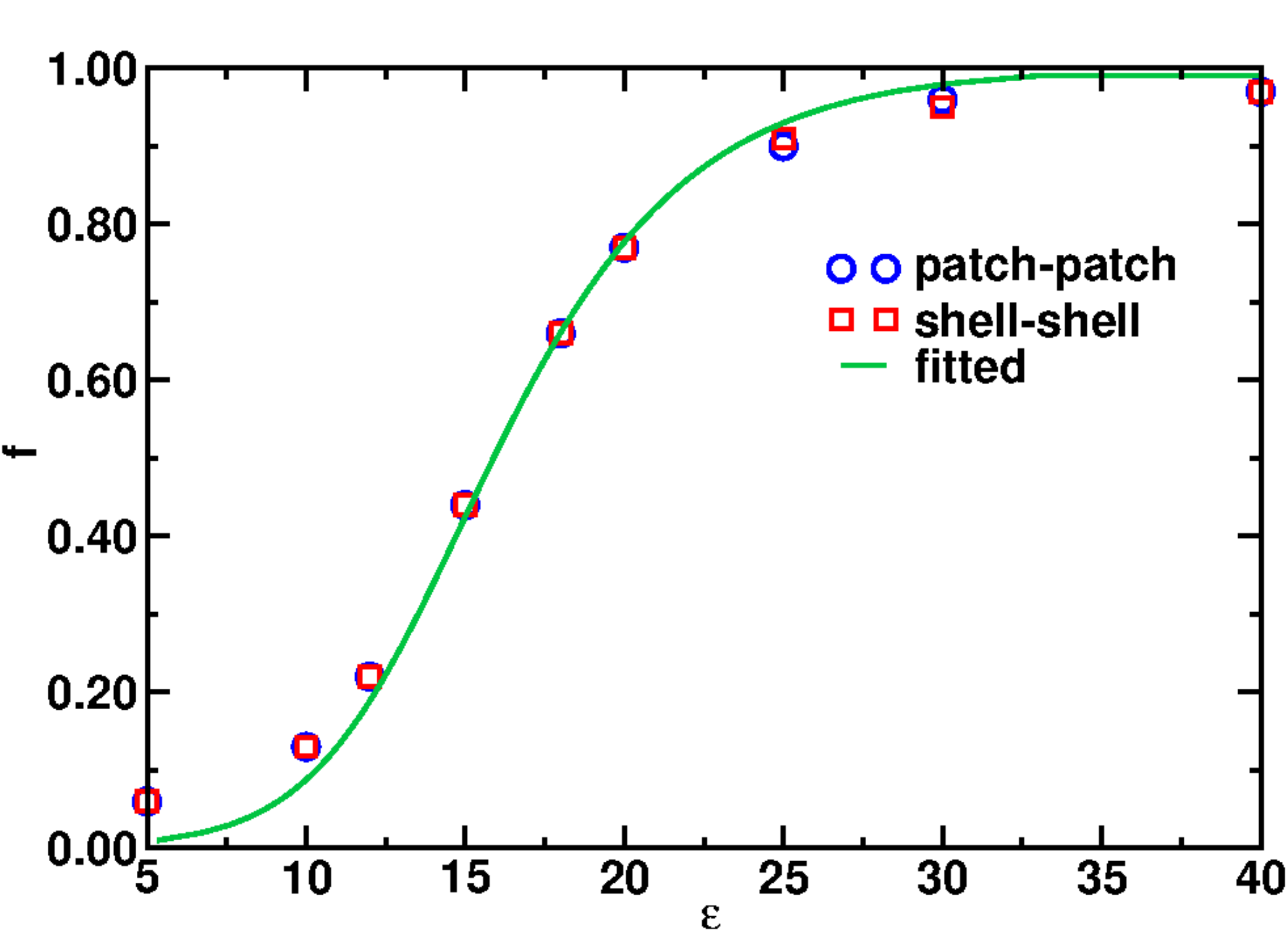}
\caption{Dimer formation and melting  for single-patch particles. Fraction of dimers ($f$) is shown as a function of  $\epsilon$.  (Blue circle): Dimer fraction for  particles with pure  PP interactions. (Red square): Dimer formation of particles with  SS interactions. The curves overlap since for the single-patch particles the shell with  ($\theta_{0} = 150 ^{\circ}$) is equivalent to the  patch with ($\theta_{0} = 30 ^{\circ}$). (Green line): Theoretical meting curve given by Eq.~\ref{Theo1}.}
\label{particle5}
\end{figure}

Next, we study the dimer formation for different $\epsilon$ values where one type of interaction is present (either PP or SS). $\theta_0 = 30^{\circ}$ for PP case and $\theta_0=150 ^{\circ }$ for SS case, respectively. All other simulation conditions are same as before. Note that the SS system should behave exactly  like the PP system  since  each particle have only one patch. Figure~\ref{particle5}  shows the dimer formation/melting curves for different $\epsilon$  values. One can see that the melting curve for PP interaction and SS interaction overlap as  expected.  
A standard theoretical result for dimer-monomer coexistence can be used to fit the simulation data. At the steady state, the fraction of dimers ($f$) can be described as: 
\begin{eqnarray}
%
\frac{(1-f)^2}{f}=e^{k(\epsilon_{0}-\epsilon)},
\label{Theo1}
\end{eqnarray}

where $k\epsilon_{0} = 7.280$ is a constant and $k\epsilon$ is the depth of the potential. $k$ ($=0.520$) can be calculated from the Figure~\ref{particle3}(b). A good agreement is found between the simulation data and Equation~\ref{Theo1}. 

\bibliography{patchy}

\begin{thebibliography}{28}%
\makeatletter
\providecommand \@ifxundefined [1]{%
 \@ifx{#1\undefined}
}%
\providecommand \@ifnum [1]{%
 \ifnum #1\expandafter \@firstoftwo
 \else \expandafter \@secondoftwo
 \fi
}%
\providecommand \@ifx [1]{%
 \ifx #1\expandafter \@firstoftwo
 \else \expandafter \@secondoftwo
 \fi
}%
\providecommand \natexlab [1]{#1}%
\providecommand \enquote  [1]{``#1''}%
\providecommand \bibnamefont  [1]{#1}%
\providecommand \bibfnamefont [1]{#1}%
\providecommand \citenamefont [1]{#1}%
\providecommand \href@noop [0]{\@secondoftwo}%
\providecommand \href [0]{\begingroup \@sanitize@url \@href}%
\providecommand \@href[1]{\@@startlink{#1}\@@href}%
\providecommand \@@href[1]{\endgroup#1\@@endlink}%
\providecommand \@sanitize@url [0]{\catcode `\\12\catcode `\$12\catcode
  `\&12\catcode `\#12\catcode `\^12\catcode `\_12\catcode `\%12\relax}%
\providecommand \@@startlink[1]{}%
\providecommand \@@endlink[0]{}%
\providecommand \url  [0]{\begingroup\@sanitize@url \@url }%
\providecommand \@url [1]{\endgroup\@href {#1}{\urlprefix }}%
\providecommand \urlprefix  [0]{URL }%
\providecommand \Eprint [0]{\href }%
\providecommand \doibase [0]{http://dx.doi.org/}%
\providecommand \selectlanguage [0]{\@gobble}%
\providecommand \bibinfo  [0]{\@secondoftwo}%
\providecommand \bibfield  [0]{\@secondoftwo}%
\providecommand \translation [1]{[#1]}%
\providecommand \BibitemOpen [0]{}%
\providecommand \bibitemStop [0]{}%
\providecommand \bibitemNoStop [0]{.\EOS\space}%
\providecommand \EOS [0]{\spacefactor3000\relax}%
\providecommand \BibitemShut  [1]{\csname bibitem#1\endcsname}%
\let\auto@bib@innerbib\@empty
\bibitem [{\citenamefont {Zhang}\ and\ \citenamefont {Glotzer}(2004)}]{Glo04}%
  \BibitemOpen
  \bibfield  {author} {\bibinfo {author} {\bibfnamefont {Z.~L.}\ \bibnamefont
  {Zhang}}\ and\ \bibinfo {author} {\bibfnamefont {S.~C.}\ \bibnamefont
  {Glotzer}},\ }\href {\doibase 10.1021/nl0493500} {\bibfield  {journal}
  {\bibinfo  {journal} {Nano Lett.}\ }\textbf {\bibinfo {volume} {4}},\
  \bibinfo {pages} {1407} (\bibinfo {year} {2004})}\BibitemShut {NoStop}%
\bibitem [{\citenamefont {Wilber}\ \emph {et~al.}(2007)\citenamefont {Wilber},
  \citenamefont {Doye}, \citenamefont {Louis}, \citenamefont {Noya},
  \citenamefont {Miller},\ and\ \citenamefont {Wong}}]{Patchy_cluster}%
  \BibitemOpen
  \bibfield  {author} {\bibinfo {author} {\bibfnamefont {A.~W.}\ \bibnamefont
  {Wilber}}, \bibinfo {author} {\bibfnamefont {J.~P.~K.}\ \bibnamefont {Doye}},
  \bibinfo {author} {\bibfnamefont {A.~A.}\ \bibnamefont {Louis}}, \bibinfo
  {author} {\bibfnamefont {E.~G.}\ \bibnamefont {Noya}}, \bibinfo {author}
  {\bibfnamefont {M.~A.}\ \bibnamefont {Miller}}, \ and\ \bibinfo {author}
  {\bibfnamefont {P.}~\bibnamefont {Wong}},\ }\href {\doibase
  10.1063/1.2759922} {\bibfield  {journal} {\bibinfo  {journal} {J. Chem.
  Phys.}\ }\textbf {\bibinfo {volume} {127}},\ \bibinfo {pages} {085106}
  (\bibinfo {year} {2007})}\BibitemShut {NoStop}%
\bibitem [{\citenamefont {Kern}\ and\ \citenamefont {Frenkel}(2003)}]{Kern}%
  \BibitemOpen
  \bibfield  {author} {\bibinfo {author} {\bibfnamefont {N.}~\bibnamefont
  {Kern}}\ and\ \bibinfo {author} {\bibfnamefont {D.}~\bibnamefont {Frenkel}},\
  }\href {\doibase 10.1063/1.1569473} {\bibfield  {journal} {\bibinfo
  {journal} {J. Chem. Phys.}\ }\textbf {\bibinfo {volume} {118}},\ \bibinfo
  {pages} {9882} (\bibinfo {year} {2003})}\BibitemShut {NoStop}%
\bibitem [{\citenamefont {Smallenburg}\ and\ \citenamefont
  {Sciortino}(2013)}]{Scior_Nat_phys}%
  \BibitemOpen
  \bibfield  {author} {\bibinfo {author} {\bibfnamefont {F.}~\bibnamefont
  {Smallenburg}}\ and\ \bibinfo {author} {\bibfnamefont {F.}~\bibnamefont
  {Sciortino}},\ }\href {\doibase 10.1038/NPHYS2693} {\bibfield  {journal}
  {\bibinfo  {journal} {Nature Phys.}\ }\textbf {\bibinfo {volume} {9}},\
  \bibinfo {pages} {554} (\bibinfo {year} {2013})}\BibitemShut {NoStop}%
\bibitem [{\citenamefont {Pawar}\ and\ \citenamefont
  {Kretzschmar}(2016)}]{Patchy_fab}%
  \BibitemOpen
  \bibfield  {author} {\bibinfo {author} {\bibfnamefont {A.~B.}\ \bibnamefont
  {Pawar}}\ and\ \bibinfo {author} {\bibfnamefont {I.}~\bibnamefont
  {Kretzschmar}},\ }\href {\doibase 10.1002/marc.200900614} {\bibfield
  {journal} {\bibinfo  {journal} {Macromol. Rapid Commun.}\ }\textbf {\bibinfo
  {volume} {31}},\ \bibinfo {pages} {150} (\bibinfo {year} {2016})}\BibitemShut
  {NoStop}%
\bibitem [{\citenamefont {Manoharan}, \citenamefont {Elsesser},\ and\
  \citenamefont {Pine}(2003)}]{Pine_patchy_2003}%
  \BibitemOpen
  \bibfield  {author} {\bibinfo {author} {\bibfnamefont {V.}~\bibnamefont
  {Manoharan}}, \bibinfo {author} {\bibfnamefont {M.}~\bibnamefont {Elsesser}},
  \ and\ \bibinfo {author} {\bibfnamefont {D.}~\bibnamefont {Pine}},\ }\href
  {\doibase 10.1126/science.1086189} {\bibfield  {journal} {\bibinfo  {journal}
  {Science}\ }\textbf {\bibinfo {volume} {301}},\ \bibinfo {pages} {483}
  (\bibinfo {year} {2003})}\BibitemShut {NoStop}%
\bibitem [{\citenamefont {Nykypanchuk}\ \emph {et~al.}(2008)\citenamefont
  {Nykypanchuk}, \citenamefont {Maye}, \citenamefont {van~der Lelie},\ and\
  \citenamefont {Gang}}]{Gang_DNA}%
  \BibitemOpen
  \bibfield  {author} {\bibinfo {author} {\bibfnamefont {D.}~\bibnamefont
  {Nykypanchuk}}, \bibinfo {author} {\bibfnamefont {M.~M.}\ \bibnamefont
  {Maye}}, \bibinfo {author} {\bibfnamefont {D.}~\bibnamefont {van~der Lelie}},
  \ and\ \bibinfo {author} {\bibfnamefont {O.}~\bibnamefont {Gang}},\ }\href
  {http://dx.doi.org/10.1038/nature06560} {\bibfield  {journal} {\bibinfo
  {journal} {Nature}\ }\textbf {\bibinfo {volume} {451}},\ \bibinfo {pages}
  {549} (\bibinfo {year} {2008})}\BibitemShut {NoStop}%
\bibitem [{\citenamefont {Park}\ \emph {et~al.}(2008)\citenamefont {Park},
  \citenamefont {Lytton-Jean}, \citenamefont {Lee}, \citenamefont {Weigand},
  \citenamefont {Schatz},\ and\ \citenamefont {Mirkin}}]{Mirkin_DNA}%
  \BibitemOpen
  \bibfield  {author} {\bibinfo {author} {\bibfnamefont {S.~Y.}\ \bibnamefont
  {Park}}, \bibinfo {author} {\bibfnamefont {A.~K.~R.}\ \bibnamefont
  {Lytton-Jean}}, \bibinfo {author} {\bibfnamefont {B.}~\bibnamefont {Lee}},
  \bibinfo {author} {\bibfnamefont {S.}~\bibnamefont {Weigand}}, \bibinfo
  {author} {\bibfnamefont {G.~C.}\ \bibnamefont {Schatz}}, \ and\ \bibinfo
  {author} {\bibfnamefont {C.~A.}\ \bibnamefont {Mirkin}},\ }\href
  {http://dx.doi.org/10.1038/nature06508} {\bibfield  {journal} {\bibinfo
  {journal} {Nature}\ }\textbf {\bibinfo {volume} {451}},\ \bibinfo {pages}
  {553} (\bibinfo {year} {2008})}\BibitemShut {NoStop}%
\bibitem [{\citenamefont {Martinez-Veracoechea}\ \emph
  {et~al.}(2011)\citenamefont {Martinez-Veracoechea}, \citenamefont {Mladek},
  \citenamefont {Tkachenko},\ and\ \citenamefont {Frenkel}}]{Francisco}%
  \BibitemOpen
  \bibfield  {author} {\bibinfo {author} {\bibfnamefont {F.~J.}\ \bibnamefont
  {Martinez-Veracoechea}}, \bibinfo {author} {\bibfnamefont {B.~M.}\
  \bibnamefont {Mladek}}, \bibinfo {author} {\bibfnamefont {A.~V.}\
  \bibnamefont {Tkachenko}}, \ and\ \bibinfo {author} {\bibfnamefont
  {D.}~\bibnamefont {Frenkel}},\ }\href {\doibase
  10.1103/PhysRevLett.107.045902} {\bibfield  {journal} {\bibinfo  {journal}
  {Phys. Rev. Lett.}\ }\textbf {\bibinfo {volume} {107}},\ \bibinfo {pages}
  {045902} (\bibinfo {year} {2011})}\BibitemShut {NoStop}%
\bibitem [{\citenamefont {Macfarlane}\ \emph {et~al.}(2011)\citenamefont
  {Macfarlane}, \citenamefont {Lee}, \citenamefont {Jones}, \citenamefont
  {Harris}, \citenamefont {Schatz},\ and\ \citenamefont {Mirkin}}]{Macfalane}%
  \BibitemOpen
  \bibfield  {author} {\bibinfo {author} {\bibfnamefont {R.~J.}\ \bibnamefont
  {Macfarlane}}, \bibinfo {author} {\bibfnamefont {B.}~\bibnamefont {Lee}},
  \bibinfo {author} {\bibfnamefont {M.~R.}\ \bibnamefont {Jones}}, \bibinfo
  {author} {\bibfnamefont {N.}~\bibnamefont {Harris}}, \bibinfo {author}
  {\bibfnamefont {G.~C.}\ \bibnamefont {Schatz}}, \ and\ \bibinfo {author}
  {\bibfnamefont {C.~A.}\ \bibnamefont {Mirkin}},\ }\href {\doibase
  10.1126/science.1210493} {\bibfield  {journal} {\bibinfo  {journal}
  {Science}\ }\textbf {\bibinfo {volume} {334}},\ \bibinfo {pages} {204}
  (\bibinfo {year} {2011})}\BibitemShut {NoStop}%
\bibitem [{\citenamefont {Wang}\ \emph {et~al.}(2015)\citenamefont {Wang},
  \citenamefont {Wang}, \citenamefont {Zheng}, \citenamefont {Ducrot},
  \citenamefont {Yodh}, \citenamefont {Weck},\ and\ \citenamefont
  {Pine}}]{Pine2015}%
  \BibitemOpen
  \bibfield  {author} {\bibinfo {author} {\bibfnamefont {Y.}~\bibnamefont
  {Wang}}, \bibinfo {author} {\bibfnamefont {Y.}~\bibnamefont {Wang}}, \bibinfo
  {author} {\bibfnamefont {X.}~\bibnamefont {Zheng}}, \bibinfo {author}
  {\bibfnamefont {E.}~\bibnamefont {Ducrot}}, \bibinfo {author} {\bibfnamefont
  {J.~S.}\ \bibnamefont {Yodh}}, \bibinfo {author} {\bibfnamefont
  {M.}~\bibnamefont {Weck}}, \ and\ \bibinfo {author} {\bibfnamefont {D.~J.}\
  \bibnamefont {Pine}},\ }\href {http://dx.doi.org/10.1038/ncomms8253}
  {\bibfield  {journal} {\bibinfo  {journal} {Nat. Commun.}\ }\textbf {\bibinfo
  {volume} {6}},\ \bibinfo {pages} {7253} (\bibinfo {year} {2015})}\BibitemShut
  {NoStop}%
\bibitem [{\citenamefont {Ho}, \citenamefont {Chan},\ and\ \citenamefont
  {Soukoulis}(1990)}]{bandgap1}%
  \BibitemOpen
  \bibfield  {author} {\bibinfo {author} {\bibfnamefont {K.}~\bibnamefont
  {Ho}}, \bibinfo {author} {\bibfnamefont {C.}~\bibnamefont {Chan}}, \ and\
  \bibinfo {author} {\bibfnamefont {C.}~\bibnamefont {Soukoulis}},\ }\href
  {\doibase 10.1103/PhysRevLett.65.3152} {\bibfield  {journal} {\bibinfo
  {journal} {Phys. Rev. Lett.}\ }\textbf {\bibinfo {volume} {65}},\ \bibinfo
  {pages} {3152} (\bibinfo {year} {1990})}\BibitemShut {NoStop}%
\bibitem [{\citenamefont {Yablonovitch}, \citenamefont {Gmitter},\ and\
  \citenamefont {Leung}(1991)}]{bandgap2}%
  \BibitemOpen
  \bibfield  {author} {\bibinfo {author} {\bibfnamefont {E.}~\bibnamefont
  {Yablonovitch}}, \bibinfo {author} {\bibfnamefont {T.}~\bibnamefont
  {Gmitter}}, \ and\ \bibinfo {author} {\bibfnamefont {K.}~\bibnamefont
  {Leung}},\ }\href {\doibase 10.1103/PhysRevLett.67.2295} {\bibfield
  {journal} {\bibinfo  {journal} {Phys. Rev. Lett.}\ }\textbf {\bibinfo
  {volume} {67}},\ \bibinfo {pages} {2295} (\bibinfo {year}
  {1991})}\BibitemShut {NoStop}%
\bibitem [{\citenamefont {Romano}, \citenamefont {Sanz},\ and\ \citenamefont
  {Sciortino}(2010)}]{phase_diagram_patchy}%
  \BibitemOpen
  \bibfield  {author} {\bibinfo {author} {\bibfnamefont {F.}~\bibnamefont
  {Romano}}, \bibinfo {author} {\bibfnamefont {E.}~\bibnamefont {Sanz}}, \ and\
  \bibinfo {author} {\bibfnamefont {F.}~\bibnamefont {Sciortino}},\ }\href
  {\doibase 10.1063/1.3393777} {\bibfield  {journal} {\bibinfo  {journal} {J.
  Chem. Phys.}\ }\textbf {\bibinfo {volume} {132}},\ \bibinfo {pages} {184501}
  (\bibinfo {year} {2010})}\BibitemShut {NoStop}%
\bibitem [{\citenamefont {Zhang}\ \emph {et~al.}(2005)\citenamefont {Zhang},
  \citenamefont {Keys}, \citenamefont {Chen},\ and\ \citenamefont
  {Glotzer}}]{Glo_diamond}%
  \BibitemOpen
  \bibfield  {author} {\bibinfo {author} {\bibfnamefont {Z.~L.}\ \bibnamefont
  {Zhang}}, \bibinfo {author} {\bibfnamefont {A.~S.}\ \bibnamefont {Keys}},
  \bibinfo {author} {\bibfnamefont {T.}~\bibnamefont {Chen}}, \ and\ \bibinfo
  {author} {\bibfnamefont {S.~C.}\ \bibnamefont {Glotzer}},\ }\href {\doibase
  10.1021/la0513611} {\bibfield  {journal} {\bibinfo  {journal} {Langmuir}\
  }\textbf {\bibinfo {volume} {21}},\ \bibinfo {pages} {11547} (\bibinfo {year}
  {2005})}\BibitemShut {NoStop}%
\bibitem [{\citenamefont {Vasilyev}, \citenamefont {Klumov},\ and\
  \citenamefont {Tkachenko}(2013)}]{Vasilyev_1}%
  \BibitemOpen
  \bibfield  {author} {\bibinfo {author} {\bibfnamefont {O.~A.}\ \bibnamefont
  {Vasilyev}}, \bibinfo {author} {\bibfnamefont {B.~A.}\ \bibnamefont
  {Klumov}}, \ and\ \bibinfo {author} {\bibfnamefont {A.~V.}\ \bibnamefont
  {Tkachenko}},\ }\href {\doibase 10.1103/PhysRevE.88.012302} {\bibfield
  {journal} {\bibinfo  {journal} {Phys. Rev. E}\ }\textbf {\bibinfo {volume}
  {88}},\ \bibinfo {pages} {012302} (\bibinfo {year} {2013})}\BibitemShut
  {NoStop}%
\bibitem [{\citenamefont {Vasilyev}, \citenamefont {Klumov},\ and\
  \citenamefont {Tkachenko}(2015)}]{Vasilyev_chrom}%
  \BibitemOpen
  \bibfield  {author} {\bibinfo {author} {\bibfnamefont {O.}~\bibnamefont
  {Vasilyev}}, \bibinfo {author} {\bibfnamefont {B.}~\bibnamefont {Klumov}}, \
  and\ \bibinfo {author} {\bibfnamefont {A.}~\bibnamefont {Tkachenko}},\ }\href
  {\doibase 10.1103/PhysRevE.92.012308} {\bibfield  {journal} {\bibinfo
  {journal} {Phys. Rev. E}\ }\textbf {\bibinfo {volume} {92}},\ \bibinfo
  {pages} {012308} (\bibinfo {year} {2015})}\BibitemShut {NoStop}%
\bibitem [{\citenamefont {Liu}\ \emph {et~al.}(2016)\citenamefont {Liu},
  \citenamefont {Tagawa}, \citenamefont {Xin}, \citenamefont {Wang},
  \citenamefont {Emamy}, \citenamefont {Li}, \citenamefont {Yager},
  \citenamefont {Starr}, \citenamefont {Tkachenko},\ and\ \citenamefont
  {Gang}}]{diamond_oleg}%
  \BibitemOpen
  \bibfield  {author} {\bibinfo {author} {\bibfnamefont {W.}~\bibnamefont
  {Liu}}, \bibinfo {author} {\bibfnamefont {M.}~\bibnamefont {Tagawa}},
  \bibinfo {author} {\bibfnamefont {H.~L.}\ \bibnamefont {Xin}}, \bibinfo
  {author} {\bibfnamefont {T.}~\bibnamefont {Wang}}, \bibinfo {author}
  {\bibfnamefont {H.}~\bibnamefont {Emamy}}, \bibinfo {author} {\bibfnamefont
  {H.}~\bibnamefont {Li}}, \bibinfo {author} {\bibfnamefont {K.~G.}\
  \bibnamefont {Yager}}, \bibinfo {author} {\bibfnamefont {F.~W.}\ \bibnamefont
  {Starr}}, \bibinfo {author} {\bibfnamefont {A.~V.}\ \bibnamefont
  {Tkachenko}}, \ and\ \bibinfo {author} {\bibfnamefont {O.}~\bibnamefont
  {Gang}},\ }\href {\doibase 10.1126/science.aad2080} {\bibfield  {journal}
  {\bibinfo  {journal} {Science}\ }\textbf {\bibinfo {volume} {351}},\ \bibinfo
  {pages} {582} (\bibinfo {year} {2016})}\BibitemShut {NoStop}%
\bibitem [{\citenamefont {Romano}\ and\ \citenamefont
  {Sciortino}(2012)}]{Sciort_Ncomm2012}%
  \BibitemOpen
  \bibfield  {author} {\bibinfo {author} {\bibfnamefont {F.}~\bibnamefont
  {Romano}}\ and\ \bibinfo {author} {\bibfnamefont {F.}~\bibnamefont
  {Sciortino}},\ }\href {\doibase 10.1038/ncomms1968
  https://www.nature.com/articles/ncomms1968#supplementary-information}
  {\bibfield  {journal} {\bibinfo  {journal} {Nature Communications}\ }\textbf
  {\bibinfo {volume} {3}},\ \bibinfo {pages} {975} (\bibinfo {year}
  {2012})}\BibitemShut {NoStop}%
\bibitem [{\citenamefont {Wang}\ \emph {et~al.}(2012)\citenamefont {Wang},
  \citenamefont {Wang}, \citenamefont {Breed}, \citenamefont {Manoharan},
  \citenamefont {Feng}, \citenamefont {Hollingsworth}, \citenamefont {Weck},\
  and\ \citenamefont {Pine}}]{Pine_2012}%
  \BibitemOpen
  \bibfield  {author} {\bibinfo {author} {\bibfnamefont {Y.}~\bibnamefont
  {Wang}}, \bibinfo {author} {\bibfnamefont {Y.}~\bibnamefont {Wang}}, \bibinfo
  {author} {\bibfnamefont {D.}~\bibnamefont {Breed}}, \bibinfo {author}
  {\bibfnamefont {V.}~\bibnamefont {Manoharan}}, \bibinfo {author}
  {\bibfnamefont {L.}~\bibnamefont {Feng}}, \bibinfo {author} {\bibfnamefont
  {A.}~\bibnamefont {Hollingsworth}}, \bibinfo {author} {\bibfnamefont
  {M.}~\bibnamefont {Weck}}, \ and\ \bibinfo {author} {\bibfnamefont
  {D.}~\bibnamefont {Pine}},\ }\href {\doibase 10.1038/nature11564} {\bibfield
  {journal} {\bibinfo  {journal} {Nature}\ }\textbf {\bibinfo {volume} {491}},\
  \bibinfo {pages} {51} (\bibinfo {year} {2012})}\BibitemShut {NoStop}%
\bibitem [{\citenamefont {Halverson}\ and\ \citenamefont
  {Tkachenko}(2013)}]{John_PRE2013}%
  \BibitemOpen
  \bibfield  {author} {\bibinfo {author} {\bibfnamefont {J.~D.}\ \bibnamefont
  {Halverson}}\ and\ \bibinfo {author} {\bibfnamefont {A.~V.}\ \bibnamefont
  {Tkachenko}},\ }\href {\doibase 10.1103/PhysRevE.87.062310} {\bibfield
  {journal} {\bibinfo  {journal} {Phys. Rev. E}\ }\textbf {\bibinfo {volume}
  {87}},\ \bibinfo {pages} {062310} (\bibinfo {year} {2013})}\BibitemShut
  {NoStop}%
\bibitem [{\citenamefont {Zheng}\ \emph {et~al.}(2016)\citenamefont {Zheng},
  \citenamefont {Wang}, \citenamefont {Wang}, \citenamefont {Pine},\ and\
  \citenamefont {Weck}}]{Pine_patch_shell}%
  \BibitemOpen
  \bibfield  {author} {\bibinfo {author} {\bibfnamefont {X.}~\bibnamefont
  {Zheng}}, \bibinfo {author} {\bibfnamefont {Y.}~\bibnamefont {Wang}},
  \bibinfo {author} {\bibfnamefont {Y.}~\bibnamefont {Wang}}, \bibinfo {author}
  {\bibfnamefont {D.}~\bibnamefont {Pine}}, \ and\ \bibinfo {author}
  {\bibfnamefont {M.}~\bibnamefont {Weck}},\ }\href {\doibase
  10.1021/acs.chemmater.6b01313} {\bibfield  {journal} {\bibinfo  {journal}
  {Chem. Mater.}\ }\textbf {\bibinfo {volume} {28}},\ \bibinfo {pages} {3984}
  (\bibinfo {year} {2016})}\BibitemShut {NoStop}%
\bibitem [{\citenamefont {Richardson}, \citenamefont {Bj{\"o}rnmalm},\ and\
  \citenamefont {Caruso}(2015)}]{Caruso-2015}%
  \BibitemOpen
  \bibfield  {author} {\bibinfo {author} {\bibfnamefont {J.~J.}\ \bibnamefont
  {Richardson}}, \bibinfo {author} {\bibfnamefont {M.}~\bibnamefont
  {Bj{\"o}rnmalm}}, \ and\ \bibinfo {author} {\bibfnamefont {F.}~\bibnamefont
  {Caruso}},\ }\href {\doibase 10.1126/science.aaa2491} {\bibfield  {journal}
  {\bibinfo  {journal} {Science}\ }\textbf {\bibinfo {volume} {348}},\ \bibinfo
  {pages} {6233} (\bibinfo {year} {2015})}\BibitemShut {NoStop}%
\bibitem [{\citenamefont {Richardson}\ \emph {et~al.}(2016)\citenamefont
  {Richardson}, \citenamefont {Cui}, \citenamefont {Bj{\"o}rnmalm},
  \citenamefont {Braunger}, \citenamefont {Ejima},\ and\ \citenamefont
  {Caruso}}]{Caruso-2016}%
  \BibitemOpen
  \bibfield  {author} {\bibinfo {author} {\bibfnamefont {J.~J.}\ \bibnamefont
  {Richardson}}, \bibinfo {author} {\bibfnamefont {J.}~\bibnamefont {Cui}},
  \bibinfo {author} {\bibfnamefont {M.}~\bibnamefont {Bj{\"o}rnmalm}}, \bibinfo
  {author} {\bibfnamefont {J.~A.}\ \bibnamefont {Braunger}}, \bibinfo {author}
  {\bibfnamefont {H.}~\bibnamefont {Ejima}}, \ and\ \bibinfo {author}
  {\bibfnamefont {F.}~\bibnamefont {Caruso}},\ }\href {\doibase
  10.1021/acs.chemrev.6b00627} {\bibfield  {journal} {\bibinfo  {journal}
  {Chem. Rev.}\ }\textbf {\bibinfo {volume} {116}},\ \bibinfo {pages} {14828}
  (\bibinfo {year} {2016})}\BibitemShut {NoStop}%
\bibitem [{\citenamefont {Xiao}\ \emph {et~al.}(2016)\citenamefont {Xiao},
  \citenamefont {Pagliaro}, \citenamefont {Xu},\ and\ \citenamefont
  {Liu}}]{Bin-2016}%
  \BibitemOpen
  \bibfield  {author} {\bibinfo {author} {\bibfnamefont {F.-X.}\ \bibnamefont
  {Xiao}}, \bibinfo {author} {\bibfnamefont {M.}~\bibnamefont {Pagliaro}},
  \bibinfo {author} {\bibfnamefont {Y.-J.}\ \bibnamefont {Xu}}, \ and\ \bibinfo
  {author} {\bibfnamefont {B.}~\bibnamefont {Liu}},\ }\href {\doibase
  10.1039/C5CS00781J} {\bibfield  {journal} {\bibinfo  {journal} {Chem. Soc.
  Rev.}\ }\textbf {\bibinfo {volume} {45}},\ \bibinfo {pages} {3088} (\bibinfo
  {year} {2016})}\BibitemShut {NoStop}%
\bibitem [{\citenamefont {Wang}\ \emph {et~al.}(2017)\citenamefont {Wang},
  \citenamefont {Jenkins}, \citenamefont {McGinley}, \citenamefont {Sinno},\
  and\ \citenamefont {Crocker}}]{ddiamond_crocker}%
  \BibitemOpen
  \bibfield  {author} {\bibinfo {author} {\bibfnamefont {Y.}~\bibnamefont
  {Wang}}, \bibinfo {author} {\bibfnamefont {I.~C.}\ \bibnamefont {Jenkins}},
  \bibinfo {author} {\bibfnamefont {J.~T.}\ \bibnamefont {McGinley}}, \bibinfo
  {author} {\bibfnamefont {T.}~\bibnamefont {Sinno}}, \ and\ \bibinfo {author}
  {\bibfnamefont {J.~C.}\ \bibnamefont {Crocker}},\ }\href {\doibase
  doi:10.1038/ncomms14173} {\bibfield  {journal} {\bibinfo  {journal} {Nat.
  Commun.}\ }\textbf {\bibinfo {volume} {8}},\ \bibinfo {pages} {14173}
  (\bibinfo {year} {2017})}\BibitemShut {NoStop}%
\bibitem [{\citenamefont {Miller}\ and\ \citenamefont
  {Cacciuto}(2009)}]{Caccuiuto-2009}%
  \BibitemOpen
  \bibfield  {author} {\bibinfo {author} {\bibfnamefont {W.~L.}\ \bibnamefont
  {Miller}}\ and\ \bibinfo {author} {\bibfnamefont {A.}~\bibnamefont
  {Cacciuto}},\ }\href {\doibase 10.1103/PhysRevE.80.021404} {\bibfield
  {journal} {\bibinfo  {journal} {Phys. Rev. E}\ }\textbf {\bibinfo {volume}
  {80}},\ \bibinfo {pages} {021404} (\bibinfo {year} {2009})}\BibitemShut
  {NoStop}%
\bibitem [{\citenamefont {Mahynski}\ \emph {et~al.}(2016)\citenamefont
  {Mahynski}, \citenamefont {Rovigatti}, \citenamefont {Likos},\ and\
  \citenamefont {Panagiotopoulos}}]{Panagiotopoulos-2016}%
  \BibitemOpen
  \bibfield  {author} {\bibinfo {author} {\bibfnamefont {N.~A.}\ \bibnamefont
  {Mahynski}}, \bibinfo {author} {\bibfnamefont {L.}~\bibnamefont {Rovigatti}},
  \bibinfo {author} {\bibfnamefont {C.~N.}\ \bibnamefont {Likos}}, \ and\
  \bibinfo {author} {\bibfnamefont {A.~Z.}\ \bibnamefont {Panagiotopoulos}},\
  }\href {\doibase 10.1021/acsnano.6b01854} {\bibfield  {journal} {\bibinfo
  {journal} {ACS Nano}\ }\textbf {\bibinfo {volume} {10}},\ \bibinfo {pages}
  {5459} (\bibinfo {year} {2016})}\BibitemShut {NoStop}%
\end{thebibliography}%

\end{document}